\documentstyle[12pt,aaspp4]{article}
\begin{document}

\title{Geometrically Thin Disk Accreting Into a Black Hole}

\author{N. Afshordi \& B. Paczy\'nski}
\affil{Princeton University Observatory, Princeton, NJ 08544--1001, USA}
\affil{E-mail: afshordi@astro.princeton.edu}
\affil{E-mail: bp@astro.princeton.edu}

\begin{abstract}
A numerical model of a steady state, thin accretion disk with a constant
effective speed of sound
is presented.  We demonstrate that `zero torque' inner boundary condition
is a reasonable approximation provided that the disk thickness, including 
the thickness of the torquing magnetic fields, is small everywhere.  It 
is likely that this conclusion is correct also for non-steady disks,
as long as the total thickness at the sonic point, $ H_c $, is much 
smaller than the radius there, $ r_c \approx r_{ms} $.  The very
existence of thin disks is not proved or disproved in this work, but
such disks are believed to exist for moderate accretion rates. 
Within our model there is a small torque at $ r_{ms} $, which
may increase disk luminosity by several percent.  An important result 
of our analysis is that the 
physically acceptable steady state solutions in our toy model exist only 
for $\alpha < 0.14 \, (100 v_s/c)^{1/3}$. 

A significant torque may be applied to a thin disk if there is a large
scale magnetic field, like in a modified Blandford-Znajek mechanism.
\end{abstract}


\section{Introduction}

Theory of accretion disks is several decades old.
With time ever more sophisticated and more diverse 
models of accretion onto black holes have been introduced.  However, when 
it comes to modeling disk spectra, conventional steady state, geometrically
thin disk models are still used, adopting the classical `no torque' inner 
boundary condition at the marginally stable orbit at $ r_{ms} $
(e.g. Blaes, Hubeny, Agol, \& Krolik 2001).  Recently, the `no torque'
condition for geometrically thin disks has been challenged by several authors 
(Krolik 1999, Gammie 1999, Agol \& Krolik 2000, to be referred as AGK.)
One of us (BP) did not agree with
their claim, and presented simple arguments why the `no torque' inner 
boundary condition is natural if an accretion disk is geometrically 
thin (Paczy\'nski 2000), but the referee could not be convinced.  Thanks 
to the electronic preprint server BP's paper is readily accessible to
all interested readers, who can judge its validity.

In this paper we present a more detailed quantitative analysis of the
inner boundary condition, refining arguments given by BP, and making them
more precise.  We find that AGK were qualitatively correct: a classical
thin steady state disk must have some torque at the $ r_{ms} $, but the
effect is not as strong as claimed by them.  Following a short historical
review of accretion disk theory in Section 2, we discuss the concept of
thin disks, and we present equations that describe a disk with a fixed value
of the effective speed of sound in Sec. 3.  In Sec. 4 the solutions of 
these equations and their topologies are studied numerically.  In Sec. 5 we
describe an analytical method to identify solutions of different nature. 
In Sec. 6 we discuss some of the physical results of our work as well
as those of Gammie (1999).  In particular, we explain why the model
and the reasoning proposed by Gammie, while bringing up an important issue,
were quantitatively incorrect.  Finally, Sec. 7 concludes this paper. 

\section{Historical outline}

A Newtonian theory of geometrically thin accretion disks was developed in 
classical papers by Pringle \& Rees (1972), Shakura \& Sunyaev (1973) and 
Lynden-Bell \& Pringle (1974), generalized for a relativistic case by
Novikov \& Thorne (1973), and reviewed by Pringle (1981).  A relativistic
effect: the bending of light trajectories leading to direct illumination of
the disk by itself, was first elaborated and calculated by Cunningham (1975,
1976).  This is marginally important for a disk accreting onto a Schwarzschild
black hole, but it is important in the Kerr case, introducing significant
non-local effects to the energy balance at small radii.
In these early
papers a geometrically thin disk accreting onto a black hole had the inner 
boundary at the marginally stable orbit, $ r_{ms} $, and the gas was freely 
falling on a tight spiral inwards of $ r_{ms} $.

There was some confusion about the nature of gas flow near the inner disk 
edge (Stoeger 1976).  The issue was first understood as a transition
from sub-sonic to supersonic flow for geometrically thick disks
(Abramowicz et al. 1978), and later for geometrically thin disks (Muchotrzeb
\& Paczy\'nski 1982).  Small $ \alpha $-parameter used in these papers
resulted in saddle-type critical points.  Matsumoto et al. (1984) found 
that for moderately large values of $ \alpha $ the critical point
was nodal-type.  These papers, and many more, were reviewed by
Abramowicz \& Kato (1989), who presented a general discussion of the
transonic flows in accretion disks.  In particular, they pointed out that
while such a transition is necessary for the existence of a steady state
accretion, it does not guarantee that a sensible global solution exists
(cf. their Fig. 2).

All the papers written in 1970s and 1980s adopted some form of `alpha'
viscosity, as there was no quantitative physical model for the effective
transport of angular momentum and the dissipation of energy within accretion
disks.  Perhaps the best qualitative description of these processes was
given by Galeyev et al. (1979).  A major breakthrough was made by
Balbus \& Hawley (1991) and Hawley \& Balbus (1991), who rediscovered a
powerful magneto-rotational instability in weakly magnetized disks, and
pointed out its relevance to accretion.  A flood of papers followed,
with powerful computers making it possible to model time dependent accretion
flows in 2-D (Armitage 1998, Stone et al. 1999, Agol et al. 2001, Stone \& 
Pringle 2001, and references therein) and in 3-D (Flemming et al. 2000, 
Hawley et al. 2001, Armitage 2001, Armitage et al.  2001, Reynolds \& 
Armitage 2001, Reynolds et al. 2001, Sano \& Inutsuka 2001, Krolik \&
Hawley 2002, Hawley \& Krolik 2002, Igumenshchev et al. 2003, and 
references therein).

A lot of semi-analytical and numerical work was done for the ADAF and CDAF
models (cf. Ball, Narayan \& Quataert 2001, and references therein).  It 
appears that at very high and at very
low accretion rates disks are geometrically thick, while thin disks may exist
only for moderate accretion rates.  If this view is correct then the early
work on geometrically thin accretion disks
may be relevant for $ 0.01 < L/L_{Edd} < 0.1 $.

A very interesting model of a geometrically thin disk with a geometrically
thick magnetic corona was recently presented by Merloni (2003).
Recent reviews of various modes of disk accretion onto black holes were
provided by Merloni (2002) and by Blaes (2002).

Yet another way to modify the picture was proposed by Blandford and Znajek
(1977), who pointed out that a large scale magnetic field may thread a black
hole, and it may extract its rotational energy.  In recent years the
original picture was modified, making it more likely that the energy 
extracted from a spinning black hole is transferred to the inner parts of
the disk, rather than directly to a distant load (Agol \& Krolik 2000, 
Li 2000a,b, Wang et al. 2002, and references therein).  Recent XMM 
observations of an AGN named MCG-6-30-15 were claimed to support this 
possibility (Wilms et al. 2001, Merloni \& Fabian 2003).
Cao \& Xu (2002) investigated
local structure of the accretion flow near the sonic point assuming
there is a steady torque exerted at the inner disk edge.

\section{A Thin Disk}

The modern 2-D and 3-D numerical simulations provided the first quantitative
and meaningful insight into the actual physics of angular momentum
transport in accretion disks.  As
computer power increases and the codes become more sophisticated a steady
progress is to be expected.  However, at this time there are significant
limitations.  The numerical
accretion flows are relatively thick, the outer boundary
is not very far out, the cooling processes are not included, the time
integrations can be carried out for a rather modest multiple of the dynamical
time scale.  Hence, when it comes to modeling disk spectra, conventional
steady state, geometrically thin disk models are still commonly used
adopting the classical `no torque' inner boundary condition at $ r_{ms} $
(e.g. Blaes, Hubeny, Agol, \& Krolik 2001).
The current ADAF or CDAF or variants of Blandford - Znajek
models do not exceed a toy model level.  This is both, natural and useful,
and the quantitative understanding will be improved with time.  For the
same reason it is useful to have a good quantitative understanding of
semi-analytical models of geometrically thin accretion disks.  The simple
toy models will remain useful even when the future 3-D numerical approach 
provides full quantitative comparison with the future observations.

At the toy model level one makes a simplifying assumption in order to
construct geometrically thin disk, to bypass the issue of thermal
(or in general internal) energy balance.  Krolik (1999) and Gammie (1999)
assumed that
the relative disk thickness is constant and small, $ H/r = const \ll 1 $,
even inwards of the marginally stable orbit, in the so called `plunging
region'.  We use an alternative assumption: the effective speed of sound
is constant and small: $ v_s/c = const \ll 1 $.  The two assumptions are
roughly equivalent as a small effective speed of sound implies small disk
thickness, and vice versa, as long as the flow is in a hydrostatic
equilibrium in the `vertical', i.e. `z' direction.  This holds even
in the supersonic flow, as long as radial velocity is much smaller than
rotation velocity, i.e. as long as the accretion time scale is much
longer than dynamical time scale.

In our view the essence of a thin disk concept is the assumption that it is
OK to integrate disk structure over its thickness, and to consider all
important variables to be functions of radius only.  In particular, thin
disk radiates away energy dissipated locally, and
no advection is allowed.  In effect a diversity of local physical
processes can be ignored, and the conservation laws of mass,
angular momentum and energy determine everything.  A simplifying
ad hoc assumption, be it $ H/r = const \ll 1 $, or $ v_c / c = const \ll 1 $,
is added in order to follow a transition from a subsonic flow at large
radii to supersonic infall at small radii.

The argument presented by Paczy\'nski (2000) in favor of applicability of
a no torque inner boundary condition was very simple.  Geometrically thin,
steady state disk accretion onto a black hole was sub-sonic at $ r > r_{in} $,
and supersonic at $ r < r_{in} $, with the inner disk radius $ r_{in} $ 
located near the marginally stable orbit, at $ r_{ms} $.  The specific angular
momentum at $ r_{in} $ is $ l_{in} $, and it satisfies the equation
$$
{ l_{in} - l_0 \over l_{in} } \approx \alpha { H_{in} \over r_{in} } \ll 1 ,
\hskip 1.0 cm {\rm if} \hskip 0.5cm \alpha \ll 1 \hskip 0.5cm {\rm and}
\hskip 0.5cm { H_{in} \over r_{in} } \ll 1
\eqno(1)
$$
(cf. eq. 4 in Paczy\'nski 2000), where $ l_0 \approx l_{in} \approx l_{ms} $
is the angular momentum integration constant, $ H $ is the disk thickness, 
and $ \alpha $ is the viscosity parameter.  Eq. (1) follows from angular 
momentum conservation in a steady state accretion flow.  Current 3-D 
numerical calculations for thick disks and tori indicate 
$ \alpha \approx 0.1 $, with large fluctuations of all physical quantities.
The inequality $ ( l_{in} - l_0 ) / l_{in} \ll 1 $ is equivalent to the
so called `no torque' inner boundary condition, or more precisely to a 
very small torque at the
inner disk edge.  This conclusion remains valid even for $ \alpha \sim 1 $,
as long as $ H_{in} / r_{in} \ll 1 $, i.e. as long as the disk remains
geometrically thin at the critical point.  Note that eq. (1) is local, i.e.
it does not matter if the parameter $ \alpha $ is constant throughout the
flow, or does it vary with radius.  The two essential assumptions are that
the disk is {\it geometrically thin}, and that the 
{\it accretion is steady state}.

Please note: we {\it do not use } eq. (1) in our numerical model calculation,
in which we make no
assumption about a relation between $ l_{in} $, $ l_{ms} $ and $ l_0 $.  The
value of the angular momentum constant is calculated adopting two assumptions:
the disk is `Keplerian' at large radii, and the flow passes smoothly through
the effective sonic point.  These two conditions select a unique value of
$ l_0 $, which turns out to be very close to $ l_{ms} $.  Also, the effective
sonic radius $ r_c $ is found to be close to the marginally stable orbit,
$ r_{ms} $.

What can be wrong with eq. (1)?  There are several possibilities.  Thin
disks, with $ H_{in}/r_{in} \ll 1 $ may be physically impossible.  We do
not know if this is true or not, as we have neither theoretical nor
observational proof either way.  Steady state accretion may not exist
in nature.  Current 3-D simulations have indicated that strict steady
state flow is not possible, but there is a possibility that when
averaged over moderate time scale a quasi steady state may still be
a sensible approximation.  The eq. (1) may hold as long as the fluctuating
disk thickness remains small at all time.  This
domain is not accessible to numerical calculations so far.  Next, there
is a global problem pointed out by Abramowicz \& Kato (1989): an accretion
flow may pass through a sonic point near $ r_{ms} $ but it may not continue
all the way into a black hole, as shown schematically in their Fig. 2.
In fact, we could not find any reference from 1980s demonstrating that
the supersonic infall for $ r < r_{ms} $ continues all the way.  This
was assumed to be obvious and the global flow pattern was not verified
numerically.  

There is an issue of terminology: what does it mean
that a disk is geometrically thin?  In the eq. (1) the disk thickness $ H $
refers to the structure which carries stresses.  If the stresses which
make the accretion possible are magnetic, as seems to be very likely, then
$ H $ must refer to the geometrical thickness of the magnetic field structure,
not just the gas layer.  The notion that the magnetic thickness may be
larger than gas thickness dates back at least to Galeyev et al. (1979),
and has found some support in recent 2-D and 3-D simulations.  Even more
importantly, all Blandford - Znajek type models are based on large scale
magnetic structures to transfer momentum and energy.  The distinction
between the magnetic and gas disk thickness was not considered at all
in the classical papers about disk accretion, and it was not mentioned by
Paczy\'nski (2000).  Yet, it is the scale height of the stresses that
is important for eq. (1).  Therefore, we emphasize that the term:
`geometrically thin disk' refers to the thickness of magnetic
structures responsible for the momentum transfer.

Another source of ambiguity comes with the terms `pressure' and $ \alpha $.
There are many different ways in which the parameter $ \alpha $ has been
defined in the literature. In the original work of Shakura \& Sunyaev
(1973), $\alpha$ was defined as the ratio of tangential stress to
the pressure of the viscous fluid. Balbus \& Hawley (1991), recognized
that this stress is in fact dominated by the magnetic terms. Early
simulations seemed to indicate that the gas pressure dominates the 
magnetic pressure, at least outside the marginally stable orbit.
Recent 3-D simulations (e.g. Hawley \& Krolik 2002)
show that magnetic pressure may dominate the gas pressure inside 
the marginally stable orbit.  Therefore, we define $ \alpha $ in terms
of total pressure, including magnetic, so it cannot exceed $ \sim 1 $.
It is not essential for our model to have constant value of $ \alpha $,
this is just the simplest assumption.

Another ambiguity is caused by the `speed of sound'. It is conventionally
defined as the speed at which gas pressure disturbances propagate, or
it may also include the effects of radiation pressure.  If the magnetic
pressure dominates then the Alfv\'en speed becomes much larger than the speed
of sound.  Rather than study the complicated effects of different speeds
at which different disturbances may propagate we introduce the
`effective speed of sound' $ v_s = (P/ \rho )^{1/2} $, where $ P $ is the
total pressure, which is relevant for the $ v_s $.
With highly tangled magnetic fields this seems to be a reasonable, though
perhaps non-orthodox definition.  It is the effective speed of sound
(the fast magnetosonic speed if magnetic pressure dominates)
that defines the transition from a sub-sonic to a supersonic flow.
It is beyond the scope of this paper to provide a rigorous justification
of our approach, but we think this simplification is sensible.  
For a disk to remain geometrically thin it is essential that the
effective speed of sound is much smaller than the speed of light,
but it does not have to be constant.  A theory which would
allow us to calculate the effective speed of sound from the first 
principles does not exist.  Therefore, we consider the simplest possible 
disk model, and we assume that throughout the disk the `effective
speed of sound' remains constant, i.e. $ v_s = const $.  This 
assumption is not essential, it is just the simplest.

We should note that a key ingredient in the above arguments, 
and in the analysis of the following sections, is the assumption 
of a hydrostatic equilibrium in the direction perpendicular to the 
thin disk.   It holds where the accretion time scale is longer
than dynamical time scale, i.e. where radial velocity is much smaller
than rotation velocity.  The `vertical' hydrostatic equilibrium
holds even while the radial flow is supersonic, approximately as long
as $ r > 2 r_g $.  The consequence of `vertical' hydrostatic equilibrium
is a simple relation between the effective speed of sound $ v_s $, rotation
velocity $ v_{rot} $, disk thickness $ H $ and the radius $ r $:
$ v_s / v_{rot} \approx H / r $.  It follows that if the disk is assumed
to be thin, i.e. $ H/r \ll 1 $, then the effective speed of sound 
must be much smaller than the speed of light, i.e. $ v_s / c \ll 1 $.
Therefore, the assumption that the disk is thin implies that magnetic
energy is assumed to be efficiently dissipated.

Following all these definitions
we present a simple thin accretion disk model in order
to demonstrate that there is no problem of the type envisioned by
Abramowicz \& Kato (1989), i.e. that once the infall becomes supersonic 
it never `turns back'.  Our simple numerical model reproduces the basic 
features of classical thin disks, and demonstrates that the transonic
solution may remain unique not only for a saddle, but also for a nodal
critical point.  But first we recall some concepts of the thin disk
models of several decades ago.  For the derivation of all equations the
reader may consult the paper by Abramowicz \& Kato (1989).

The single most important difference between Newtonian accretion and an
accretion onto a black hole is the presence of a marginally stable orbit
in the latter.  Newtonian gravity varies strictly as a power of radius,
therefore binding energy and the `Keplerian' angular momentum vary as power
laws of radius, and this makes self-similarity an acceptable simplifying 
assumption for some accretion flows.  This is not
the case in general relativity: at small radii `Keplerian' quantities are no
longer power laws of radius, they do not even vary monotonically.  The
'Keplerian' angular momentum and the corresponding binding energy reach a
minimum at $ r_{ms} $, the radius of a marginally stable orbit.  This
feature is reproduced with a pseudo-Newtonian potential (Paczy\'nski
\& Wiita 1980):
$$
\Psi = - { GM \over r - r_g }, \hskip 2.0cm r_g = { 2GM \over c^2 } . 
\eqno(2)
$$
For a test particle on a `Keplerian' circular orbit we have a relation
$$
{ d \Psi \over dr } = 
{ GM \over \left( r - r_g \right) ^2 } = { v_K^2 \over r } ,
\eqno(3)
$$
where $ v_K $ is the `Keplerian' rotational velocity.  It follows that 
`Keplerian' angular momentum, $ l_K $, angular velocity, $ \Omega _K $, 
and binding energy, $ e_K $, are given as
$$
l_K = v_K r = \left( GMr \right) ^{1/2} \left( { r \over r - r_g } \right) ,
\hskip 1.3cm \Omega _K = { v_K \over r } =
\left( { GM \over r^3 } \right) ^{1/2} \left( { r \over r - r_g } \right) ,
\eqno(4a)
$$
$$
e_K = \Psi + { v_K^2 \over 2 } = 
- ~ { GM \left( r - 2 r_g \right) \over 2 \left( r - r_g \right) ^2 } 
\eqno(4b)
$$
At very large radii, $ r/r_g \gg 1 $, the eqs. (2-4) asymptotically become
Newtonian, but the differences are large at small radii.  In particular, a
minimum value of `Keplerian' angular momentum and binding energy is reached at
$ r = r_{ms} = 3 r_g $, just as it does in the relativistic Schwarzschild case.
The orbits with $ r > r_{ms} $ are stable, while those with $ r < r_{ms} $
are unstable, with a marginal stability at $ r_{ms} $.

Let us consider the simplest disk model with zero pressure.  In this case all
streamlines are identical with particle trajectories, i.e. they are nested
`Keplerian' orbits.  This structure can be extended to small radii, but all
orbits inwards of $ r_{ms} $ are unstable: a small perturbation makes a
particle spiral into a black hole.  Therefore, a zero pressure disk
is truncated at $ r_{ms} $, it cannot extend inwards of $ r_{ms} $.  Such
a disk has zero geometrical thickness, and it is static, i.e. there is no
accretion.

To make disk accretion possible it is necessary to introduce some means of
angular momentum exchange between nearby streamlines.  In general, this
will require a non-zero total pressure $ P $, a non-zero effective
speed of sound
$ v_s \approx \left( P/ \rho \right) ^{1/2} $, and a finite disk thickness
$ H $.  For simplicity we assume the effective speed of sound to be constant:
$$
{ H \over r } \approx { v_s \over v_K } \ll 1 , \hskip 0.8cm
v_s = const, \hskip 0.8cm
{ v_s^2 \over c^2 } \approx { P \over \rho c^2 } \ll 1 , \hskip 0.8cm
\Sigma = \int \rho dz ,
\eqno(5a)
$$
We remind the reader that pressure $ P $ is the total pressure: 
$$
P \equiv P_{tot} = P_{gas} + P_{rad} + P_{mag} + P_{turb} .
\eqno(5b)
$$
The effective speed of sound $ v_s $ becomes close to
Alfv\'en speed when the magnetic pressure dominates.

We assume that the stresses responsible for angular momentum 
transport are proportional to the total pressure and the shear rate.
The torque exerted by these stresses across radius $r$ is
$$
g= -\alpha ~ 2 \pi r^2 ~ \Omega^{-1}_K ~ \left( { d\Omega \over d\ln r } 
\right)
~ \int P dz =
-\alpha ~ 2 \pi r^2 ~ \left( { d\ln\Omega \over d\ln r } \right)
\left({\Omega \over \Omega_K } \right)  
~ v_s^2 \Sigma .
\eqno (6)
$$
The dimensionless factor $\alpha$ is assumed to be less than
unity, and for simplicity we take it to be constant.
Again, just as it was the case with the `effective speed of sound',
the assumption that $ \alpha $ is constant is not essential,
it is just the simplest.
Note, that $ \Omega $ varies monotonically with radius, i.e.
the inner disk rotates faster, and angular momentum is transported outwards.

The $ \alpha P $ term, no matter how small, redistributes angular momentum 
within the disk, and accretion becomes possible.  In a steady state each 
disk element gradually loses its angular momentum, 
and slowly spirals inwards.  For $ r > r_{ms} $ the process can 
be envisioned as a motion along ever smaller nearly `Keplerian'
orbits, with gradually decreasing angular momentum.  This brings
matter close to $ r_{ms} $.  Following any additional loss of angular momentum
the gas cannot find any `Keplerian' orbit and must plunge into the black hole
along a spiral, approximately conserving angular momentum.  

Let us assume a steady state, with the mass accretion rate 
$ \dot M = const < 0 $, and the radial flow velocity $ v_r < 0 $ assumed
to be a function of radius only.  The conservation laws of mass and angular
momentum are
$$
\dot M = 2 \pi r \Sigma v_r , \hskip 2.0cm
g = \left( - \dot M \right) \left( l - l_0 \right) , 
\eqno(7)
$$
where $ l_0 $ is the integration constant (cf. Abramowicz \& Kato 1989).
Combining eqs. (6) and (7) we obtain
$$
\left( - v_r \right) \left( l - l_0 \right) =
\alpha r v_s^2 \left( - { d \ln \Omega \over d \ln r } \right) 
\left( { \Omega \over \Omega _K } \right) .
\eqno(8)
$$
It is reasonable to expect that $ l_0 \approx l_{ms} $ (to be verified later),
and the last equation may be used to estimate the radial flow velocity in
the nearly `Keplerian' thin disk for $ r > r_{ms} $ and $ |v_r|/v_s \ll 1 $.

Note, that in eq. (8) only two quantities vary a lot within the
flow: $ v_r $ and $ \left( l - l_0 \right) $, while the others are either
constant or vary slowly.  In particular, according to the classical theory
of thin accretion disks, near the $ r_{ms} $ and inwards
of $ r_{ms} $ angular momentum is almost conserved.  This implies that
$$
\left( - { d \ln \Omega \over d \ln r } \right) \approx 2 , \hskip 1.0cm
{\rm therefore} \hskip 1.0cm
\left( - v_r \right) \left( l - l_0 \right) \approx 2 \alpha r v_s^2 
~ \left( { \Omega \over \Omega _K } \right) .
\eqno(9)
$$
The numerical solutions of the following sections will verify
this assumption.

We expect that when matter reaches $ r \approx r_{ms} $ any additional loss of angular 
momentum puts it on a spiral infall toward the black hole, during which
energy and angular momentum are approximately conserved, i.e. 
$$
e = - { GM \over r - r_g } + { v_r^2 \over 2 } + { v_{rot}^2 \over 2 } 
\approx e_{ms} = - ~ { c^2 \over 16 } , \hskip 2.0cm
l = v_{rot} r \approx l_{ms} = 3 \sqrt{1.5} ~ { GM \over c } ,
\eqno(10)
$$
where $ v_{rot} $ is the rotational velocity component of the infall.
This is verified with the numerical integrations in the following
sections. 

To calculate a transition from the sub-sonic radial flow
to the supersonic flow we need the equation
of motion (cf. Abramowicz \& Kato 1989)
$$
v_r { d v_r \over dr } - \left( \Omega ^2 - \Omega _K^2 \right) r
+ { v_s^2 \over \Sigma } { d \Sigma \over dr } +
v_s^2 { d \ln \Omega _K \over dr } = 0 .
\eqno(11)
$$
Differentiating the mass conservation law (eq. 7) we obtain
$$
{ v_s^2 \over \Sigma } ~ { d \Sigma \over dr } =
- {v_s^2 \over r } - v_s^2 ~ { d \ln v_r \over dr } ,
\eqno(12)
$$
and combining eqs. (11) and (12) we find
$$
\left( v_r^2 - v_s^2 \right) { d \ln v_r \over d \ln r } -
\left( \Omega ^2 - \Omega _K^2 \right) r^2 + v_s^2
\left( { d \ln \Omega _K \over d \ln r } - 1 \right) = 0.
\eqno(13)
$$

We introduce dimensionless constants and variables defined as
$$
a \equiv { v_s \over c } \ll 1 , \hskip 1.0cm
b \equiv { l_0 \over l_{ms} } \approx 1 , \hskip 1.0cm
0 < \alpha < 1 ,
\eqno(14a)
$$
$$
\omega \equiv { \Omega \over \Omega _K } , \hskip 1.0cm
v \equiv - ~ { v_r \over c } > 0 , \hskip 1.0cm
x \equiv { r \over r_g } .
\eqno(14b)
$$
The equation of motion (13) and conservation of angular momentum (9)
may be written in dimensionless form as
$$
{ d \ln v \over d \ln x } = { 1 \over \left( v^2 - a^2 \right) } ~
\left[ \left( \omega ^2 - 1 \right) ~ { x \over 2 (x-1)^2 } ~ + ~
2.5 ~ a^2 ~ { x - 0.6 \over x - 1} \right] ,
\eqno(14)
$$
$$
\omega = { 1.5 \sqrt{3} ~ b (x-1) v \over 
x \left[ x^{1/2} v - 2 \sqrt{2} ~ \alpha a^2 (x-1) \right] } ,
\eqno(15)
$$

\section{Critical Points, Physical Solutions and their Topology}

We seek a physical solution which has the properties
$$
v/a \ll 1 , \hskip 1.0cm \omega \approx 1 , \hskip 1.0cm
{\rm for } \hskip 0.5cm x \gg 3 ,
\eqno(16a)
$$
$$
v = a , \hskip 1.0cm {\rm for} \hskip 0.5cm x = x_c ,
\eqno(16b)
$$
$$
v/a \gg 1 , \hskip 1.0cm {\rm for} \hskip 0.5cm x \ll x_c .
\eqno(16c)
$$
i.e. the flow is sub-sonic and the disk is `Keplerian' at large radii, the 
flow passes through a critical point, and it becomes supersonic at small radii. 

To better understand the behavior of the physical solutions of eq. (14),
and to be able to integrate it numerically near the critical point,
we decompose it into two equations
$$
\frac{d\ln v}{dt} = \left( \omega ^2 - 1 \right) ~ { x \over 2 (x-1)^2 } ~ + ~
2.5 ~ a^2 ~ { x - 0.6 \over x - 1},
\eqno(17a)
$$
$$
\frac{d\ln x}{dt} = v^2 - a^2, 
\eqno(17b)   
$$    
which, together with eq. (15), form a two-dimensional autonomous 
dynamical system. 
The variable $t$, is a dummy variable and should not be confused with 
physical time.

As mentioned earlier, the physical solution must
pass a critical point to reach the supersonic regime, otherwise
the solution would not be single-valued as the right hand side of eq. (14)
diverges. Now, the critical points of eq. (14) are the fixed points 
of eqs. (17) and an understanding of the nature of these fixed
points is necessary for constructing a physical solution.

At any fixed point, the right hand sides of eqs. (17) vanish.
Eq. (17b) implies that fixed points are always at $v=a$, as required by
eq. (16b).  The right hand side of eq. (17a)
at the critical point, together with
eq. (15) (with $v=a$) lead to a complicated algebraic 
equation for $x$ at a given $b$, or vice versa. The analysis is 
significantly simplified for 
small values of $a$ and the corresponding relation between $b$ and $x_c$
is shown in Fig. 1. We see that, in general, there are either
no fixed points or two fixed points in the system, while at the 
critical value of $b=1+O(\alpha a)$, there is only a single degenerate fixed 
point at $x \approx 3$.

The nature of the fixed points can be investigated 
analytically by linearizing eqs. (17) around each 
of them. It turns out that the left fixed point 
(smaller value of $x$) is always saddle-like,
i.e. the two eigenvalues have opposite signs.
The right fixed point is of the spiral type 
(two complex eigenvalues) when the two fixed points 
are far apart, but it switches to the nodal type (two 
eigenvalues of the same sign) as the two fixed points get closer.
This behavior is shown in Figs. 2-5 which 
is the typical behavior of a saddle-node bifurcation
(Guckenheimer \& Holmes 1983). 

The integrations presented in Figs. 2-5 were calculated numerically,
except near the critical points, where an analytic expansion was used.
All four figures have the same disk
parameters: $ a = v_s / c = 0.01 $ and $ \alpha = 0.15 $.  They differ
in the adopted value of the angular momentum constant $ b $.  The family
of solid lines shown in each figure represents possible solutions of
the eq. (14).  The solutions of special interest for us are those which
satisfy boundary conditions (16a,b,c).  As it turns out none of these
figures has a desired solution, but the figures present the changes in
the structure of the critical points.

Fig. 2 presents a saddle critical point at $ x_c = 2.7 $, and a spiral point
at $ x_c = 3.4 $.  The Phase Portraits for $ b > 0.9997 $ (including
$ b > 1 $) is qualitatively similar to Fig. 2.  The Phase Portrait changes
when the angular momentum constant $b$ is reduced, as shown in Fig. 3,
where the two critical points are saddle at $ x_c = 2.94 $, and nodal at
$ x_c = 3.1 $.  When $b$ is reduced down to $ b = 0.9947 $ the two critical
points merge, as shown in Fig. 4.  This corresponds to the point marked
$(x_c,1)$ in Fig. 1.  When $b$ is reduced even more then there is no critical
point, as demonstrated with Fig. 5, and as was anticipated (cf. Fig. 1).

A prominent feature of Figs. 2-5 is the 
`Keplerian' solution. In the sub-sonic regime, ($v \ll a$),
the the right hand side of eq. (14) is very large for small values of $a$ 
unless $\omega \simeq 1$ (almost `Keplerian' angular velocity). 
This implies that the only slowly varying solution of eq. (14)
is an almost `Keplerian' one. The radial velocity for a `Keplerian' 
solution is obtained by setting $\omega = 1$ in eq. (15),
$$
 v_0 = \frac{2\sqrt{2}\alpha a^2 x}{x^{3/2}/(x-1)- 1.5 \sqrt{3}b}+O(a^4).
\eqno(18)
$$
Figs. 2-5 show that all the
other solutions tend to merge with the `Keplerian' solution 
as $x$ decreases. This implies that the behavior of the physical solution 
close to $r_{ms}$ is almost
independent of the boundary condition at large radii, since all 
different solutions merge as they move inwards.  Therefore, it is easy to
satisfy the boundary condition given with eq. (16a).  However, none
of the solutions shown in Figs. 2-5 satisfies all three conditions given
by eqs. (16a,b,c).
      
It is clear that a physical solution may pass through a
saddle or a nodal fixed point, but not a spiral.
Furthermore, we assume that 
a physical solution is analytic. This singles out only
two ways of passing a nodal fixed point 
(slow or fast modes), because a linear superposition
of fast and slow modes is non-analytic (when $v$
is expressed as a function of $x$). In the case of 
a saddle fixed point, only two solutions (in the direction 
of two eigenvectors) can pass the fixed point.
A conclusion that can be drawn from this consideration is 
that, in general, the `Keplerian' solution does not 
analytically pass the fixed points and only for discrete 
values of the angular momentum constant $b$ a physical solution 
is possible. 

Now, we are ready to analyze different topologies that can
occur in the parameter space of our model.  For a given value of $a$
the nature of critical points can be calculated analytically for any 
pair of values of $ \alpha $ and $x_c$, or equivalently $ \alpha $ and $b$.
Fig. 6 shows the possibilities for $ a = v_s / c = 0.01 $. The 
horizontal axis, $x$, is the position of a fixed point and the 
vertical axis is the parameter $\alpha$ that appears in the viscosity model 
(eq. 5). The fixed points are
of the saddle-type, on the left side of the dashed region, of the 
nodal type inside the dashed region, and of the spiral type on the 
right side of the dashed region. The solid curve aBCd shows the position 
of physical solutions that allow passage of the `Keplerian' curve
through the critical point.  The critical points present in Figs. 2-4 
are indicated in Fig. 6 with diamonds numbered 2, 3, 4, 3, and 2, 
correspondingly, with $ \alpha = 0.15 $ for all of them.
The diamonds with numbers 7, 9, and 11, correspond to critical points
presented in Figs. 7, 9, 11; they are all located on the solid line aBCd.

We note that for $\alpha$ smaller than some 
critical value $\alpha_{SN}$, the physical critical
point is of the saddle type (segment aB), while for larger values
of $\alpha$ it is of the nodal type (segment Bd). The value of 
$\alpha_{SN}$ depends on $a$. For example for $a=0.01$ we have
$\alpha_{SN} = 0.08$ (cf. Fig. 6). 
 
A critical point of the saddle type, shown in Figs. 7 and 8,
is the type that is usually encountered in astrophysics. The best
known examples are the solar wind (Parker 1958) and Roche lobe overflow
in binary stars. As argued above, when the parameters $ a $ and $ \alpha $
are fixed then physical solution exists only for a unique 
value of $b$ and the corresponding value of $x_c$. 
   
In the case of a nodal critical point, the passage is possible
through fast or slow \footnote{The fast/slow direction refers
to the direction of the eigenvector with the larger/smaller
eigenvalue.} directions. However, a generic solution, which
is a combination of fast and slow modes, passes the critical point
in the slow direction since the slow solution dominates the fast one
close to the critical point. In this sense, a physical solution 
that passes a nodal point in the fast direction, is unique, 
similar to the saddle critical
point. This kind of passage is seen in the BC section of the aBCd 
curve in Fig. 6 and and example of the phase portrait is shown in 
Figs. 9 and 10.
Note that, again similar to the saddle point solutions, there is a
smooth transition from the sub-sonic to the supersonic regime.

Finally, the physical solution may pass the critical point in 
the slow direction. However, as argued above, this does not fix 
the value of $b$ since it is the generic behavior of the solutions,
and only by requiring the analyticity of the physical solution one may
determine $b$ uniquely. The points on the Cd segment in Fig. 6 are of 
this type, and an example of the phase portraits is plotted in Figs. 11 
and 12.  
  
The transition from the former type of nodal passage to the latter
occurs at $\alpha_{NN}$. For example, for
$a=0.01$, $\alpha_{NN}\simeq 0.14$ (point C in Fig. 6). The curve
aBCd, at the transition point C, is tangent to the boundary of the nodal
and saddle regions, where the fast and slow directions merge. 
    
The main difference between the last type of passage and the previous two 
types is that there is a sharp change in slope, right after passing the
critical point. The reason is that the `Keplerian' curve, which is 
connected to the slow direction, turns from stable (as $x$ decreases) 
in the sub-sonic regime to unstable in the supersonic 
regime, and so the physical solution departs from the `Keplerian' 
curve after the passage. Since the slopes are significantly larger
far from the `Keplerian' curve, there is a sharp change in slope.      

While we have described some properties of solutions with a nodal
critical point, a general discussion of these matters is beyond the
scope of this paper.  The central issue for us is the inner boundary
condition for a steady state geometrically thin accretion disk.
For a given value of the effective speed of sound $ v_s = ac $,
and for $\alpha \le 0.14$ we found a unique numerical solution,
in agreement with the finding of Artemova et al. (2001).
We verified the uniqueness of numerical solution in two ways.
First, we began with a guess of the location of the critical point $ x_c $,
and we searched for the value which provided a solution close to `Keplerian'
at large radii.  This provided a unique value of the angular momentum constant
$b$ for the given $a$ and $\alpha$.  We also started numerical integrations
of the eqs. (17 and 15) at large radii, where we adopted a `Keplerian'
model.  We searched for the value of the angular momentum constant $b$
for which the solution would pass through the critical point.  There was a
unique value of $b$ and $x_c$ found for the given $a$ and $\alpha$ and it
was identical to that found with the first method.  There was no difference
in the numerical procedure for solutions which had a saddle or a nodal 
critical point.  

We found that for $ a = 0.01 $ and $ 0.01 \le \alpha \le 0.14 $ there are
unique solutions for the sub-sonic, steady state flow with the effective
speed of sound assumed to be constant.
The critical point location varied monotonically in the range 
$ 2.8286 \le x_c \le 3.1058 $
and the angular momentum constant varied monotonically in the range
$ 1.00068 \le b \le 0.99528 $, i.e. we had $ x_c \approx x_{ms} = 3.0 $ and
$ b \approx 1.0 $.  These results are not in any way affected by the
supersonic flow at $ x < x_c $, i.e. no matter how complicated is
the flow for $ x < x_c $ it has no influence on the flow
for $ x \ge x_c $.  However, there might be a problem 
of the type presented by Abramowicz \& Kato (1989)
in their Fig. 2: the flow passing through the critical point
may formally reverse at some radius $ x < x_c $, i.e. it may be globally
impossible.  This is not the case if the effective speed of sound
is assumed to be constant also in the supersonic flow: our solutions
continue smoothly all the way to the black hole, and the angular momentum
remains almost constant within the supersonic flow, as expected.

We were not able to find steady-state solutions for $ \alpha > 0.14 $.
Perhaps physical solutions cannot pass a nodal point
in the slow direction, and the accretion is non steady for $ \alpha > 0.14 $.
We shall explore the consequences of variable effective speed of sound
in section 5.2.

\section{Some Analytic Results}

In order to have an analytic understanding of the behavior of the  
solutions and the dependence on the parameters $\alpha$ and $a$ we
attempt to replace eq. (14) by a simplified version of it, which retains
the main topological features of the phase portrait, i.e. critical 
points and the `Keplerian' curve, and also the dependence on $\alpha$ and
$a$ for small $a$'s. First let us define $\nu = v/a$ and $\nu_0 =v_0/a$,
where $v_0$, defined in eq.(18), is the `Keplerian' radial velocity. Next
we make the following 
assumptions
$$
x \simeq 3, \, \, \nu_0 \simeq 1, \, \, \omega \simeq 1,
\eqno(19)
$$ 
which all follow from the assumption of $a \ll 1$ and
restricting the study to the vicinity of the `Keplerian' curve
(that includes the critical points).  After these substitutions,
being careful not to drop terms that are crucial for the main 
topological features, we end up with
$$
\frac{d\nu}{dx} \simeq -\frac{\alpha(\nu - \nu_0)}{\sqrt{6} a
\nu_0 (\nu-1)}.
\eqno(20)
$$     
We see that the `Keplerian' curve, for which $d\nu/dx$ vanishes,
and the critical points, where $\nu = \nu _0 = 1$, are preserved.

Next, we try to find a workable approximation for $\nu_0(x)$. 
To do so, we notice that the interesting transitions in the
nature of the physical solution (Fig. 6) happens where the critical
points are close to the maximum of $\nu_0(x)$, which is also close
to $1$. With these approximations, we can use eq. (18) to get
$$
\frac{d\ln \nu_0}{dx} \simeq \frac{d\nu_0}{dx} \simeq 1/3 - 2B(x-3),
\eqno(21)
$$  
where
$$
B^{-1} = 32 \sqrt{6}\alpha a,
\eqno(22)
$$  
 which can be integrated to give 
$$
\nu_0(x) \simeq 1+ B[(x_c-x_b)^2-(x-x_b)^2], \,\,
x_b = 3+ B^{-1}/6.
\eqno(23)
$$  
$x_c$ is the position of the critical point, which is related
to $b$ by setting $\omega=1$ in eq.(15), and  $x_b$ is the Saddle-Node
bifurcation point, the boundary of the Saddle region and the Nodal 
region in Fig. 6.  
   
Now, let us define the new variables $\xi$, $\Delta$ and $\mu$
$$
\xi \equiv B^{1/2}(x-x_b), \, \,
\mu \equiv B^{1/2}\sqrt{6}a/ \alpha = [\frac{\sqrt{6}
a}{32\alpha^3}]^{1/2}, \, \,
\Delta \equiv \nu-\nu_0.
\eqno(24)
$$
   
In terms of these new variables, eq.(20) becomes
$$
 \frac{d\Delta}{d\xi} =
-\frac{\mu^{-1}\Delta}{(1+\xi^2_c-\xi^2)(\Delta+\xi^2_c-\xi^2)}+2\xi.
\eqno(25)
$$
Now, it is easy to find the eigenmodes $M$ of the critical point, by
setting $\Delta$ equal to $M(\xi-\xi_c)$ and requiring that it satisfies
eq. (25) to the first order in $\xi-\xi_c$. This yields
$$
M=-\frac{1}{2\mu}(1-4\mu\xi_c \pm \sqrt{1-8\mu\xi_c}),
\eqno(26)
$$    
which already gives the three saddle, nodal and spiral 
regions in Fig. 6. 

We do not intend to elaborate any further
on the behavior of the solutions of eq.(25). This is 
mainly because the main property of the physical solution
which is connecting the asymptotically `Keplerian' curve
to the critical point, is a global property and so  
its analytical investigation is not straightforward.
The only clear conclusion that we can draw from eq.(25) is 
that the value of $\xi_c$ for which the physical solution 
exists is a function of $\mu$. The transitions in the nature
of the critical point, points B (SN) and C (NN) in
Fig. 6, that were discussed in the last section, occur at 
specific values of $\mu$ which can be already fixed 
from the numerical results in Fig. 6 and the definition 
of $\mu$ in eq. (24)
$$
\mu_{SN} \simeq 1.22, \, \,
\mu_{NN} \simeq 0.53.
\eqno(27)
$$
Of course the advantage is that now we can apply this result to
all small values of $a$ and not only $a=0.01$. This yields
$$
(\frac{a}{0.01})<(\frac{\alpha}{0.14})^3, 
\hskip 1.0cm {\rm \, Nodal\, Point\, Through\, Slow
\, Direction,} 
$$
$$
(\frac{\alpha}{0.14})^3 <(\frac{a}{0.01})<(\frac{\alpha}{0.08})^3,
\hskip 1.0cm {\rm \, Nodal
\, Point \, Through \,Fast \, Direction,}
$$
$$
(\frac{\alpha}{0.08})^3 < (\frac{a}{0.01}) , 
\hskip 1.0cm {\rm \, Saddle \, Point.}
\eqno(28)
$$ 
\noindent
The numerical values calculated for $ a = 0.02 $ agree very well with these
scalings.
    
\section{Discussion}

After numerical and analytical study of the properties of the
solutions of eq. (14) in the last two sections, we are now ready
to discuss the physical picture within the framework of our
approximations. Of course the question of accuracy of these 
approximations should be addressed as well, and we intend to do so,
at least in part.  
 
Our numerical solutions fully confirm our guesses that the angular momentum
constant $ l_0 \approx l_{ms} $ (i.e. $ b \approx 1 $), and
that $ d \log \Omega / d \log r \approx -2 $ near $ r_{ms} $ and
for $ r < r_{ms} $ (cf. eq. 9, and the text between eq. 8 and 9).
Note: we have not assumed that $ l \approx l_{ms} $, this
was obtained numerically as a consequence of two conditions: the
disk had to be nearly `Keplerian' at large radii, and the transition
through the sonic point had to be smooth (cf. eqs. 16).

\subsection{Torque at the Sonic Point}

The starting point of this
investigation was a recent controversy about the significance of 
the torque at the sonic point of a thin accretion disk.  The traditional
picture of Novikov \& Thorne (1973) argues for a `zero torque'
boundary condition at the marginally stable orbit $r_{ms}$.  Historically,
the possibility of a significant torque at the sonic point due to strong
magnetic fields was first brought up in Page \& Thorne (1974).  This
issue was pursued more seriously by Krolik (1999), Gammie(1999), 
Agol \& Krolik (2000).

Our disk model, with a smooth flow passage through the effective sonic
point, has a small but non-zero torque there.  The most relevant
way to quantify the importance of this effect is to look at the energy
generation rate due to differential rotation, which follows from the
conservation laws:
$$
{ d L_d \over dr } = g \times \left( - { d \Omega \over dr } \right) =
\left( - \dot M \right) \left( l - l_0 \right) 
\left( - { d \Omega \over dr } \right) ,
\eqno(29)
$$
(cf. eq. 7).  In the classical model of a geometrically thin disk with a no
torque at $ r_{ms} $ and `Keplerian' rotation for $ r > r_{ms} $,
eq. (29) may be integrated to obtain
$$
L_{d,0} = \int _{r_{ms}}^{\infty} { d L_d \over dr } dr =
\left( - \dot M \right) { c^2 \over 16 } =
\left( - \dot M \right) c^2 \epsilon _0 ,
\hskip 1.0cm
\epsilon _0 = { 1 \over 16 } .
\eqno(30)
$$

Eq. (29) may be evaluated numerically for our disk model, and the
results are shown in Fig. 13 for $ a = 0.01 $.
A thick line corresponds to the
classical model and two thin lines correspond to our models with
$ \alpha = 0.01 $ and $ \alpha = 0.1 $, respectively.  The differences are
small but noticeable.  The total energy released is reduced by fraction of
one percent in the disk with $ \alpha = 0.01 $ because the angular momentum
constant is slightly larger than $ l_{ms} $, with $ b = 1.00068 $.  The 
energy is increased by $ 2.5\% $ for $ \alpha = 0.1 $ model, because the
angular momentum constant is slightly reduced, with $ b = 0.99636 $. 

Next, we calculated the change of accretion efficiency:
$$
{ \Delta \epsilon \over \epsilon _0 } =
{ L_d \over \left( - \dot M \right) c^2 \epsilon _0 } ~ - ~ 1 ,
\eqno(31)
$$
where $ L_d $ was calculated with eq. (29) for 12 disk models with
all combinations of the two parameters: $ a = $ 0.01, 0.02, 0.04, and
$ \alpha = $ 0.01, 0.05, 0.10, 0.14.  The results are presented in Fig. 14
as a function of angular momentum constant: $ b = l_0 / l_{ms} $.  The
important result is that the efficiency of accretion increases with
increasing disk thickness, which is proportional to the parameter $ a $,
and with increasing parameter $ \alpha $.  
Within the range of parameters presented in
Fig. 14 the efficiency is increased by up to 11\% of $ \epsilon _0 $,
i.e. from $ 0.0625 \dot M c^2 $ to $ 0.0694 \dot M c^2 $.

There are two related effects contributing to a small change in accretion
efficiency.  The fact that angular momentum constant $ l_0 $ is not
exactly equal $ l_{ms} $ affects energy generation at all radii, even in
the `Keplerian' outer disk, as it is apparent in eq. (29).  In addition,
the disk near $ r_{ms} $ is not completely `dark' as it would be if the
torque at $ r_{ms} $ were exactly zero.

The disk scale height $ H $ at the $ r_{ms} $ may be estimated as
$$
\left( { H \over r } \right) _{ms} \approx 
\left( { a c \over v_{rot} } \right) _{ms} =
 \left( { 8 \over 3 } \right) ^{1/2} ~ a \approx 1.63 a ,
\eqno(32)
$$
i.e. $ a = 0.04 $ corresponds to $ (H/r)_{ms} \approx 0.065 $.
As we assume the effective speed of sound to be constant, the disk thickness
increases with radius, and we reach $ H/r = 1 $ at $ r \approx r_g /(2a^2) $.
The `Keplerian' condition at `large radii' makes sense only for $ a \ll 1 $,
and this limits the range of parameters for which our disk models make sense.

\subsection{Is Nodal Passage in the Slow Direction Physical?}

Inspection of Figs. 2, 3, 4, 7, 8, 9, and 10 shows that the geometry
of transonic flow appears to be similar when the critical point is
of a saddle type or a nodal type, provided the solution passes the
latter in the fast direction.  The transition from the saddle to
the nodal (fast) geometry is smooth, with no apparent problem.  
The sub-sonic flow connects well to a `Keplerian' disk at large radii,
and the supersonic flow approaches a free fall at small radii.
This trouble free region corresponds to the aBC segment (thick line)
in Fig. 6,
which ends at point C, where it contacts the borderline of the fixed 
points of the spiral type, and where $ \alpha = 0.14 $.  Obviously, no 
physical solution can pass through a spiral point.  Above point C the
analytical solutions pass the nodal critical point in the slow
direction, and these solutions do not appear to be physical (cf. Fig. 11), 
as there is a sharp change in the slope of the solution (see
the end of Sec. 4).  
We do not know if they are unstable, or perhaps no truly steady-state
flow is possible for $ \alpha > 0.14 $ (for $ a = 0.01 $).

The study of the stability of these points with the assumption of
constant $v_s$ and $\alpha$ is even less satisfactory than the steady
state solutions. Nevertheless, Kato, Honma, and Matsumoto (1988) show
that, within these assumptions, local instabilities occur when 
$$
\alpha_{SS} \Omega(r_c) > |\frac{dv}{dr}|_c,
\eqno(33)
$$    
where $\alpha_{SS}$ is the Shakura-Sunyaev value of $\alpha$,
($\alpha_{SS}=2\omega \alpha$). The dotted portion of the 
the physical curve in Fig. 7, which covers almost all of the 
sector Cd, shows the critical points that satisfy this criterion.
This indicates that the nodal passage in the slow direction 
is probably unstable. We should point out that, contrary to 
the conventional picture (originally suggested by Matsumoto et al. 1984),
this is not a generic property of the nodal critical points, but
a generic property of the passage in the slow direction of 
the nodal point, where $|dv/dr|_c$ in eq. (33) is small.  

We cannot prove that the transition through a nodal-type point
in the fast direction is the only physically acceptable solution, i.e.
that it is unique, but this seems to be likely
upon inspection of Figs. 9, 10, 11, 12, and it also agrees with
the finding of Artemova et al. (2001).  If correct, this could
resolve the ambiguity discovered by Matsumoto et al. (1984).
Unfortunately, we cannot offer a simple physical explanation 
for the transition from saddle-type to nodal-type critical
points while the parameter $ \alpha $ increases.  We stress that
the physical nature of nodal points is outside the scope of our paper,
which is concentrated on the issue of the inner boundary condition
for geometrically thin steady state accretion disks.

\subsection{Discussion of Gammie (1999) paper}

There remains the issue which gave rise to this paper: why did Krolik (1999),
Gammie (1999) and Agol \& Krolik (2000) claim that the torque at
inner boundary may be large even for thin, steady state
disks, which remain thin even in the `plunging region'?  Let us consider 
the case presented by Gammie (1999), as he provides the fullest model 
calculation.  Gammie assumed that
`... the disk is thin, $ c_s^2 / c^2 \ll 1 $...',
and `... that $ \alpha \ll 1 $  so that  magnetic fields make a negligible
contribution to the hydrostatic equilibrium of the disk ...'.  He also
writes: `This picture leads one to consider a steady, axisymmetric inflow
close to the equatorial plane of the Kerr metric.'  These assumptions
are practically identical to ours. 

Some of the assumptions we made in this paper, 
and some made by Gammie are similar: we all assume that geometrically 
thin steady state accretion disks exist, and that magnetic fields are
confined to the disk. 
Our model, with a constant and small effective speed of sound, implies
that the flow remains geometrically thin also in the supersonic region,
for $ r < r_c $.  Gammie also assumed that the flow was geometrically
thin everywhere.  In fact his flow thickness, $ H $, was decreasing for
$ r < r_{ms} $ according to $ H/r = const \ll 1 $ (cf. his eq. 4).

A disagreement appears in the treatment of the transition from the
`disk' to the `plunging region'.  Gammie adopts one set of assumptions
for the flow in the `disk', for $ r > r_{in} $, and a very different set of
assumptions for the flow in the `plunging region', for $ r < r_{in} $.
At $ r_{in} $ the model undergoes a dramatic jump in the adopted physical
conditions: it is gas dominated and has negligible magnetic field for 
$ r > r_{in} $, and it has no gas pressure and it is magnetic field dominated
for $ r < r_{in} $.  No justification is offered for this jump in physical
conditions.  The location of $ r_{in} $ is adopted to be close
to the marginally stable orbit, $ r_{in} \approx r_{ms} $.  

We do not introduce any $ r_{in} $, and we treat the `disk' and the
`plunging region' with the same equations.  We require a smooth transition
from sub-sonic flow in the `disk' to supersonic flow in the `plunging region'
through a critical (sonic)
point at some $ r_c $.  We make no assumption about the value of $ r_c $,
the condition of a smooth passage determines that value uniquely.  We find
that the critical point is located close to the marginally stable orbit, 
$ r_c \approx r_{ms} $, but it is in general not at $ r_{ms} $.  The
requirement of a smooth passage through
the sonic point imposes a stringent restriction on the solution, it acts
like a boundary condition.  

Another problem with Gammie's model is the assumption that the
flow thickness is proportional to radius, i.e. that the flow becomes
geometrically thinner in the `plunging region', while the effective speed
of sound rapidly increases with the decreasing radius in his model.  This is 
incompatible with the hydrostatic equilibrium in the direction perpendicular
to the equatorial plane.  In fact there is no reference to the hydrostatic
equilibrium in his paper.  That equilibrium should hold as long as
radial velocity remains much smaller than rotational velocity, i.e.
as long as the accretion time scale remains much longer than the dynamical
time scale.  As the effective speed of sound increases in Gammie's
model, the flow thickness
should also increase, not decrease.  A quasi-static expansion needed
to maintain hydrostatic equilibrium is likely to conserve magnetic flux,
i.e. $ B_r H $ and $ B_{\varphi} H $, where $ H $ is the disk thickness.
As $ H $ increases the magnetic field gets weaker, $ B \sim 1/H $, and 
the magnetic torque at a given radius $ r $ is reduced:
$ \sim B_r B_{\varphi} r H \sim r / H $.  

For the same reason the Alfv\'en speed $ v_A $ is reduced by flow expansion,
so as to maintain approximate hydrostatic equilibrium:
$ v_A / v_{rot} \sim H / r $.

It is beyond the scope of this discussion to speculate how the conclusions
of AKG about the torque at $ r_{in} $ would change if
the two major inconsistencies in Gammie's analysis were corrected: the
jump at $ r_{in} $ was replaced with full continuity of all physical
quantities, and the consequences of hydrostatic equilibrium in the 
`vertical' direction were incorporated in the model.  These two problems
are serious enough to doubt the conclusions: the torque at $ r_{ms} $
may be large and the accretion efficiency may exceed 100\%.

No direct comparison is possible between our model and Gammie's (1999)
model, as the latter does not specify the value of $ H/r $.  Amazingly,
his results appear to be independent of $ H/r $ value, while our results
are sensitive to the assumed disk thickness, as given with the $ a $
parameter (cf. eq. 32, and Fig. 14).

The consequences of ad hoc assumptions may be very diverse.  AGK
claim the importance of magnetic fields increases with
decreasing radius, and becomes dominant in the `plunging region'.
Using different ad hoc assumptions Li (2002) comes to the opposite
conclusion in his simple analytic model.

A large part of Agol \& Krolik (2000) analysis remains correct: there
may be a torque applied to the inner edge of the accretion disk provided
there is a large scale magnetic field, possibly threading the black hole.
This is a modified Blandford-Znajek mechanism (e.g. Li 2000, Wang et al.
2002, and references therein).  The gaseous
disk may remain geometrically thin while the stresses are transmitted 
by a large scale magnetic field which has a vertical scale height comparable
to radius.  However, this picture is conceptually very different from
the case considered by us and by Gammie (1999), as we and Gammie assume
that the magnetic fields are confined to a geometrically thin flow.

\section{Conclusions}

The condition given with the eq. (1) of this paper is local, and it is
approximately
valid even if the parameter $ \alpha $ varies with radius.  It requires
the assumption of a steady state to hold.  However, we think it is reasonable
to expect that it may also hold for a disk with an accretion which is
steady in a statistical sense only, as long as the disk thickness
does not exceed the value $ H \ll r $ throughout the fluctuations
cycle.  Obviously, it is important to verify this claim with the
full 3-D time dependent simulations.  However, current numerical
models are limited by the available computer power to disks which are 
much thicker than the one considered in this paper.  They are also 
limited by the absence of cooling
processes which are essential for the formation of thin disks.  
At this time there is no observational
or theoretical reason to exclude the possibility that geometrically thin
disks exist.  Currently, there is also no way to prove their existence,
as the relevant thickness is that of
the layer over which the magnetic fields transferring
angular momentum extend, and this quantity is not readily observable.

We have found a unique steady state solution for a simple model with
a constant effective speed of sound $ v_s = 0.01 c $, and a given viscosity 
parameter $ \alpha < 0.14 $.  The solution was found by approaching the
critical point from large radii, i.e. from the sub-sonic flow side,
where at large radii the disk was assumed to be `Keplerian'.  The same
solution was found starting integrations from the critical point and
proceeding out and seeking a solution approaching `Keplerian' at large radii.
No assumption was made about the value of the angular momentum constant 
$ l_0 $, which was determined by our choice of boundary conditions:
`Keplerian' disk at large radii, and a smooth passage through the sonic point.

Our numerical model confirms the validity of our eq. (1): we find that in
a thin steady state disk the sonic point $ r_c $ is located close to the
marginally stable orbit $ r_{ms} $, and the value of angular momentum 
constant $ l_0 $ is close to $ l_{ms} $.

The supersonic flow has no effect on the critical point,
as expected on general grounds.  In particular, we demonstrated that, for
geometrically thin inflow, with constant effective speed of sound, there is 
no global problem of the type envisioned by Abramowicz \& Kato (1989) in 
their Fig. 2. Once the flow passes the critical point it plunges into the
black hole along a spiral, roughly conserving angular momentum, as
asserted in many papers written in the 1980s.

In our disk model the torque at the sonic point is small, but not
exactly zero, and the total disk luminosity is modified, but only
by several per cent.  This small torque and a modest change
of accretion efficiency makes the largest difference
near $ r_{ms} $, where the accretion flow is no longer `dark',
as it was under strict `no torque' inner boundary condition.
This effect should be included in calculating spectra of thin disks.

We emphasize that for a disk to be thin its
thickness $ H $ must include the magnetic structures
responsible for angular momentum transfer, not just the
gas layer.  The effective speed of sound $ v_s $ is affected by the total
pressure, which includes the contribution of magnetic pressure (cf. eq. 5b).

Just as discovered by Matsumoto et al. (1984) we also found that
the critical point is of the saddle-type for small values of $ \alpha $,
and becomes a nodal-type for large $ \alpha $,
in our case for $ \alpha > 0.08 $.  Unfortunately, we cannot provide a simple 
physical explanation for this transition.  However, we presented plausible
arguments that the transition through the nodal-type critical point is
unique, as there is only one solution which appears to be smooth.
This was also demonstrated by Artemova et al. (2001).

We found that our steady state solutions exist only for $ \alpha < 0.14
(100 v_s/c)^{1/3} $,
and it appears that there are no physically sensible solutions for $ \alpha > 
\alpha _{crit} \approx 0.14 \, (100 v_s/c)^{1/3}$.
It is not clear if this is a general result or just an artifact due to
simplicity of our model.  The issue is not the particular value of
$ \alpha _{crit} $, which is certainly model dependent, but the very
existence of $ \alpha _{crit} $.  Unfortunately, we cannot offer a
simple physical explanation for this finding.

Krolik (1999), Gammie (1999) and Agol \& Krolik (2000) claimed that their
geometrically thin steady state disks had large torques at $ r_{ms} $,
and that accretion efficiency could be larger than
100\% of $ \dot M c^2 $, i.e. a thin accretion flow could extract energy
from a spinning black hole.  This may be correct if there are large scale
magnetic fields, like those proposed for
the modified Blandford-Znajek mechanism, also referred to as a magnetic
coupling model (e.g. Agol \& Krolik 2000, Li 2000, Wang et al. 2002,
and references therein).  However, if the magnetic fields are
confined to a thin accretion flow, as in the model described by Gammie (1999),
the very high accretion efficiency is almost certainly
a consequence of several inconsistent assumptions (cf. our Sec. 6.3).
It would be interesting to learn how Gammie's results change 
when his errors are corrected, but this is beyond the scope of this paper.

With our thin disk model we found that
there is a small torque at $ r_{ms} $, and there is a corresponding
modest change of accretion efficiency.  It is not possible to compare
our results directly with those obtained by Gammie as he
does not specify how thick his disk is.  

We are very grateful to the anonymous referee who relentlessly pressured
us to refine our analysis.
It is a great pleasure to acknowledge many useful and critical comments
by Dr. J. Goodman.


\newpage

\appendix
\section{`Quasi-adiabatic' accretion}

Our assumption that the accretion flow remains geometrically thin at all
radii: in the sub-sonic part for $ r > r_{cr} $, and in the
supersonic part for $ r < r_{cr} $, is physically equivalent to the
assumption that magnetic fields dissipate effectively at all radii.
In this Appendix we discuss the consequences of a possibility
that there is no field dissipation at radii smaller than some transition
radius $ r_{tr} $.  In such a case
the magnetic field pressure and the disk thickness may increase
for $ r < r_{tr} $.  Krolik (1999) pointed out the possibility
that the differential rotation within the supersonic flow may increase the
effective speed of sound so much that `... we expect the Alfv\'en speed
in the fluid frame to be $ \sim $ c ...' (Krolik 1999).  Obviously, if
the magnetic field dominates the flow then the effective speed of sound
is the same as Alfv\'en speed, and it also becomes relativistic.  

The largest energetically possible increase of the effective speed of sound
may be obtained by assuming a `quasi-adiabatic' flow: all energy transfered 
from differential rotation into the accretion flow is used to increase
magnetic energy, no magnetic energy is dissipated, and none is radiated away.
This is certainly a limit
which is not likely to be reached in realistic flows.
Nevertheless, we explored this possibility as a limiting case, and
we refer to it as a `quasi-adiabatic' approximation.
Note that the model developed by Gammie (1999) was also `quasi-adiabatic'.

We repeated the calculations of our models
with $ a = 0.01 $.  In the first case we adopted $ \alpha = 0.01 $,
in the second $ \alpha = 0.1 $, and we obtained:
($ x_c=2.82859$, $ b=1.00068$) and ($ x_c=3.04505$, $b=0.99636 $),
for the two cases, respectively.  Obviously, these values
were identical to the models described
in Section 4.  We continued these solutions some distance into the
supersonic part of the flow, down to a transition radius $ x_{tr} $, where
we abruptly switched the flow from having constant effective speed of
sound $ v_s = 0.01 c $, and constant $ \alpha $, to a `quasi-adiabatic' 
approximation.  All physical quantities varied continuously, but their
derivatives experienced a jump.

We assumed that at $ x_{tr} $ the ratio of magnetic 
pressure to total pressure was equal to $ \alpha $.  For $ r < r_{tr} $
the energy of differential rotation was pumped into tangled magnetic
fields, hence the magnetic pressure increased.  No magnetic energy was
dissipated, and the gas pressure was assumed to change
adiabatically, i.e. the gas pressure decreased because
of decompression.  Therefore, the ratio $ P_{mag}/P_{tot} $ increased
down the flow for $ x < x_{tr} $, and the parameter $ \alpha $ 
increased correspondingly.  We assumed that hydrostatic equilibrium 
was maintained in the direction perpendicular to the disk equatorial 
plane, as that equilibrium
was established on a fraction of rotational period, while the flow
followed many rotations prior to crossing the black hole horizon.
The gas density rapidly decreased as the flow accelerated toward
the black hole and expanded in vertical direction.  The vertical
expansion of the flow was a consequence of the increase in
the effective speed of sound, which became practically equal
to the Alfv\'en speed.

We found that if the transition from $ v_s = ac = const $
to a `quasi-adiabatic' flow was made sufficiently far into the supersonic
region, with $ x_{tr} < 2.79 $ and $ x_{tr} < 2.78 $ for the two
models, respectively, the flow proceeded smoothly all the way to the black
hole, and remained similar to the flow in our original model.  
At the transition radius, $ x_{tr} \approx 2.8 $, the flow was moderately 
supersonic, with the
effective Mach number of 1.3 and 2.4, for the two cases, respectively.
Down the flow, for $ x < x_{tr} $, the magnetic pressure increased, and
the parameter $ \alpha $ also increased to $ \alpha \approx 1 $.
The effective speed of sound gradually increased by a factor
$ \sim 3 $, increasing the flow thickness correspondingly,
but the flow remained supersonic.

When the transition was made at a larger value of $ x_{tr} $, i.e. at a 
smaller effective Mach number, the flow pattern was initially `normal', 
i.e. the $ \alpha $, the effective Mach number and the Alfv\'en speed all 
increased gradually
with decreasing radius.  However, at still smaller radii the flow 
first became sub-sonic, and later the numerical solution `reversed' the 
flow direction, changing the sign of $ v_r $, i.e. the flow proceeded back
toward large radii.  Obviously, this is physically impossible,
as pointed out by Abramowicz \& Kato (1989).
Superficially a global steady state solution could not
exist for the transition radius larger than approximately $ 2.8 r_g $.
However, it is virtually certain that before the flow would `reverse',
the deceleration would generate a shock, and a transition to a sub-sonic
accretion.  Perhaps a steady state flow
could be maintained, with the second sonic point located down
the flow from the shock location, and the matter ultimately plunging
into the black hole.  However, these complexities are beyond the scope
of our paper.

The results of our numerical experiment are easy to understand qualitatively.
As the dissipation of magnetic energy is switched off (`quasi-adiabatic' 
approximation) the magnetic pressure builds up, and the flow expands
in vertical direction.  Had the transition been made at a large radius,
the disk would become geometrically thick.  The very assumption that
the disk is thin is obviously equivalent to the assumption that the
dissipation of magnetic field is efficient.

Gammie (1999) assumed that there was no magnetic dissipation in his
model i.e. his model was also `quasi-adiabatic'.  But there were
major differences between his and our models.  Gammie assumed that 
$ H/r = const \ll 1 $ at all $ r < r_{in} \approx r_{ms} $,
while our flows thickened appreciably for $ r < r_{tr} < r_{ms} $.
Gammie assumed that magnetic fields had a very special smooth
spiral geometry, while we assumed they were tangled.  His solutions
for the flow were very different from ours.  The main result of our
experiment is that in a `quasi-adiabatic' approximation the flow thickens,
which is hardly surprising.  We also found that properties of a
`quasi-adiabatic' flow strongly depend on the value of the transition
radius $ r_{tr} = x_{tr} r_g $.  Gammie did not explore this diversity,
effectively adopting $ r_{in} = r_{tr} = r_{ms} $.`

We were curious to study the consequences of a `quasi-adiabatic'
approximation, and we were intrigued to find
that even within this approximation our solutions were very 
different from those obtained by Gammie.  However, this experiment is not
relevant to the main topic of our paper, which was focused on the
properties of thin accretion disk.

\newpage

\begin{figure}[t]
\plotfiddle{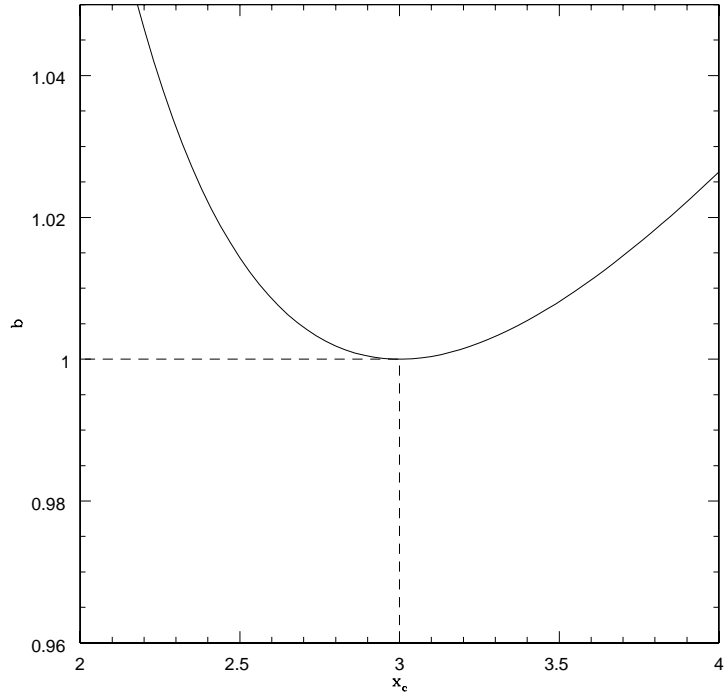}{8cm}{0}{50}{50}{-160}{-80}
\caption{The angular momentum constant $b$ is shown as a function of the 
position of the critical 
point $x_c$ for $a\rightarrow 0$. We see that, for $b>1$, there are two fixed 
points, at $b=1$, there is one degenerate fixed point at $x=3$ (marginally 
stable orbit), and for
$b<1$ no fixed point exist.}
\end{figure}

\begin{figure}[t]
\plotfiddle{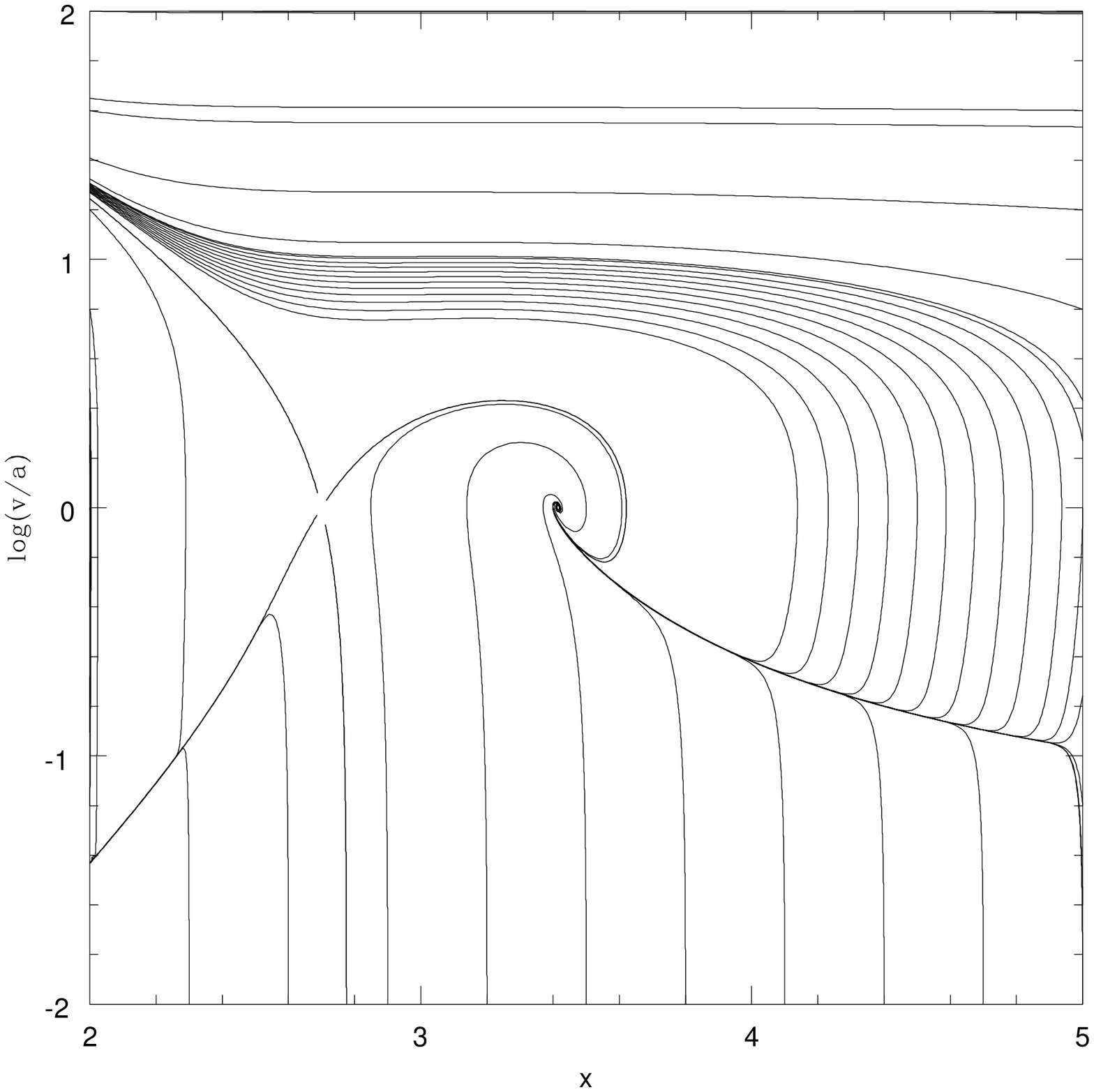}{8cm}{0}{50}{50}{-160}{-80}
\caption{The Phase Portrait for $a=0.01, \alpha = 0.15$ and $b=0.9997$.
There is a saddle point at $x=2.7$ and a spiral point at $x=3.4$.}
\vspace{1cm}
\plotfiddle{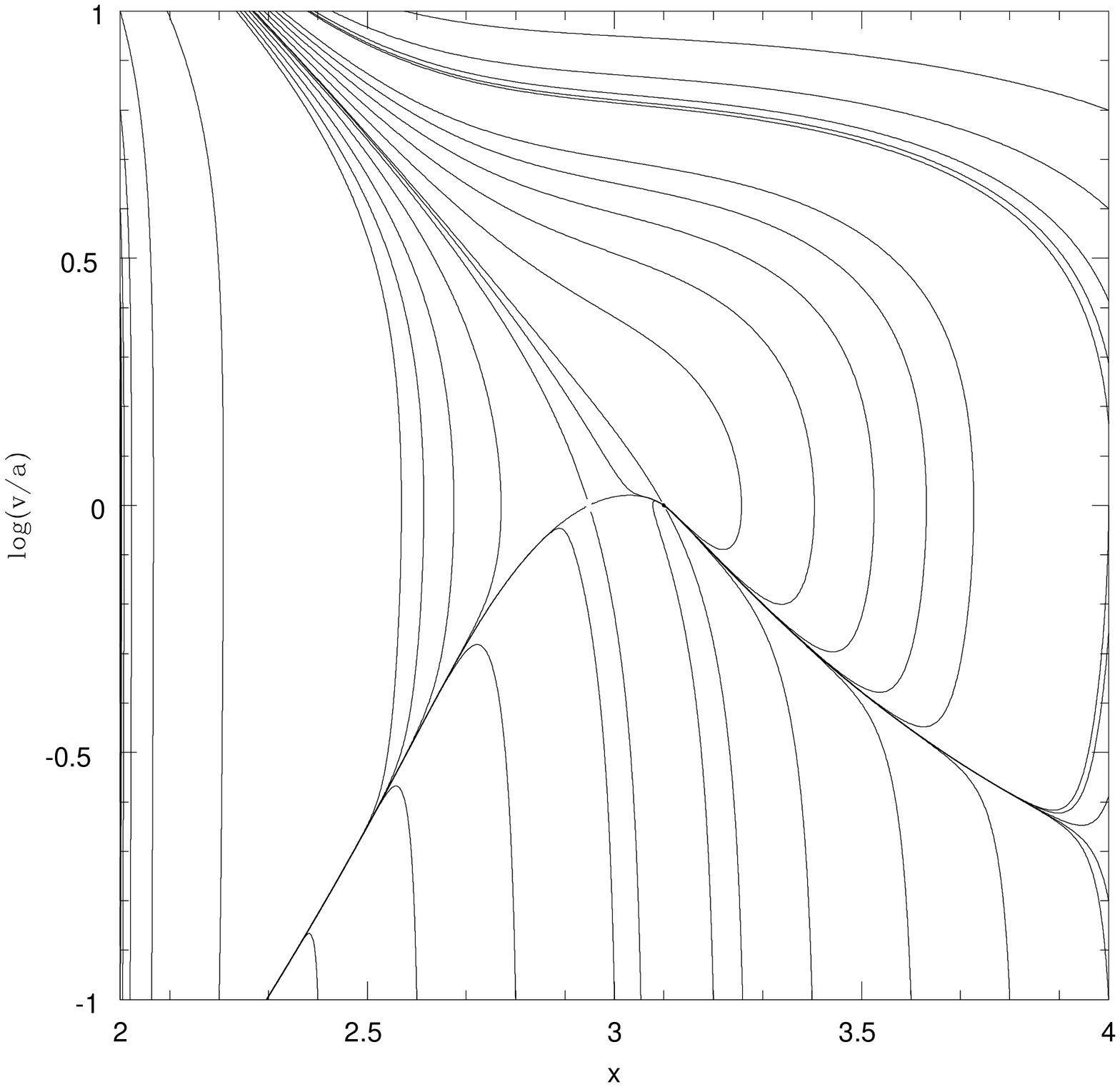}{8cm}{0}{50}{50}{-160}{-80}
\caption{The Phase Portrait for $a=0.01, \alpha = 0.15$ and $b=0.9949$.
There is a saddle point at $x=2.94$ and a nodal point at $x=3.1$.}
\end{figure}

\begin{figure}[t]
\plotfiddle{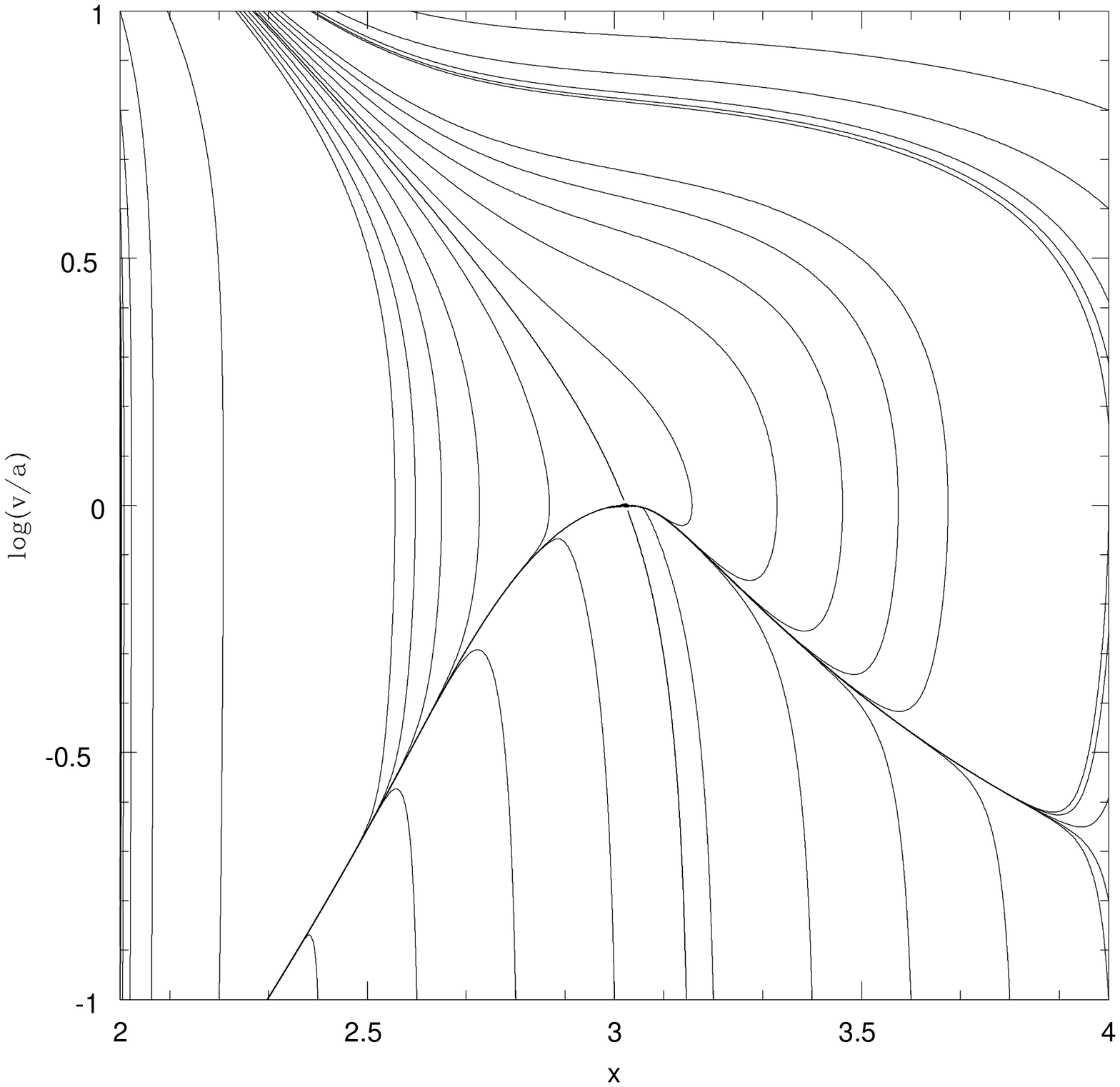}{8cm}{0}{50}{50}{-160}{-80}
\caption{The Phase Portrait for $a=0.01, \alpha = 0.15$ and $b=0.9947$.
There is a degenerate saddle-node point at $x=3.0228$. }

\vspace{1cm}
\plotfiddle{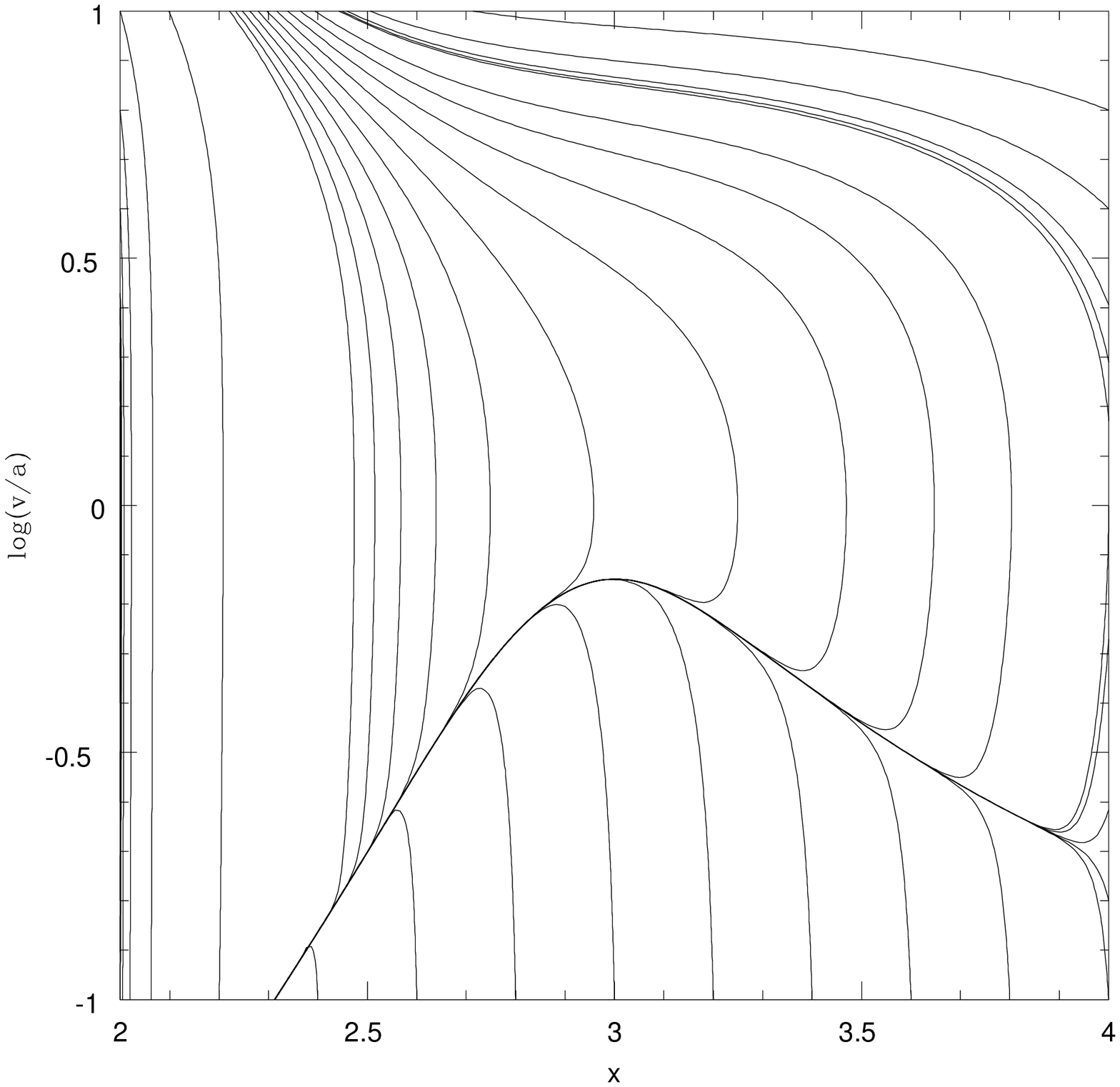}{8cm}{0}{50}{50}{-160}{-80}
\caption{The Phase Portrait for $a=0.01, \alpha = 0.15$ and $b=0.9927$.
There is no fixed point for these parameters. }
\end{figure}

\begin{figure}[t]
\plotfiddle{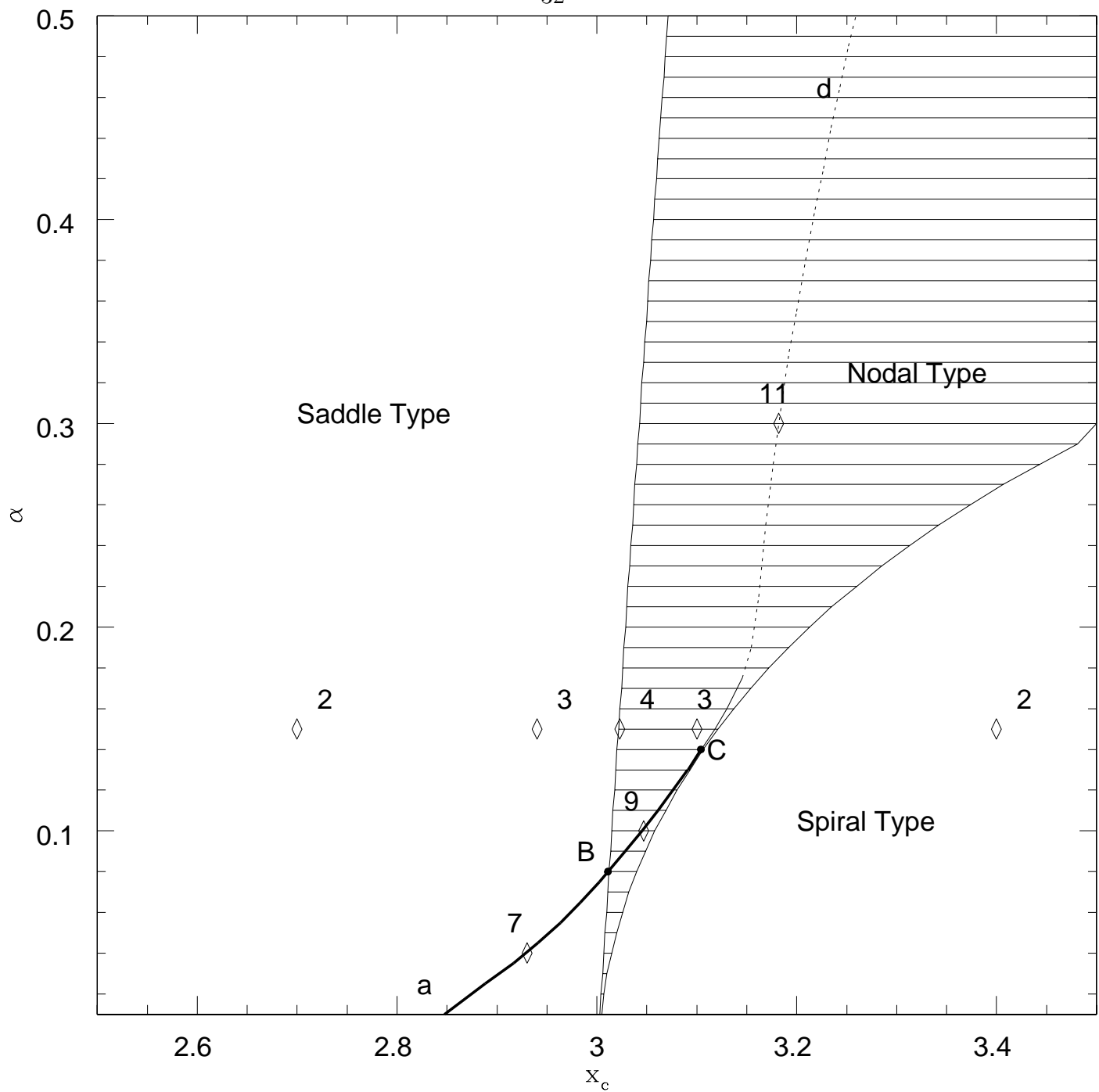}{17cm}{0}{100}{100}{-300}{-170}
\caption{The parameter space of the problem for $a=v_{s}/c =0.01$. 
$x_{c}$ is the radius of the critical point in units of $r_g$
and $\alpha$ is the viscosity parameter.  
The curve aBCd shows the position of the fixed points 
that allow physical solutions (cf. Sec. 4), while the dotted 
section of the curve are unstable solutions according to eq.(30). 
Also the parameters of phase portraits
in the other figures are represented by numbered diamonds. 
The number of each diamond is the number of the associated figure
in this paper. The Phase Portrait presented in Fig. 5 has no fixed points,
and therefore it is not represented in this figure.}
\end{figure}

\begin{figure}[t]
\plotfiddle{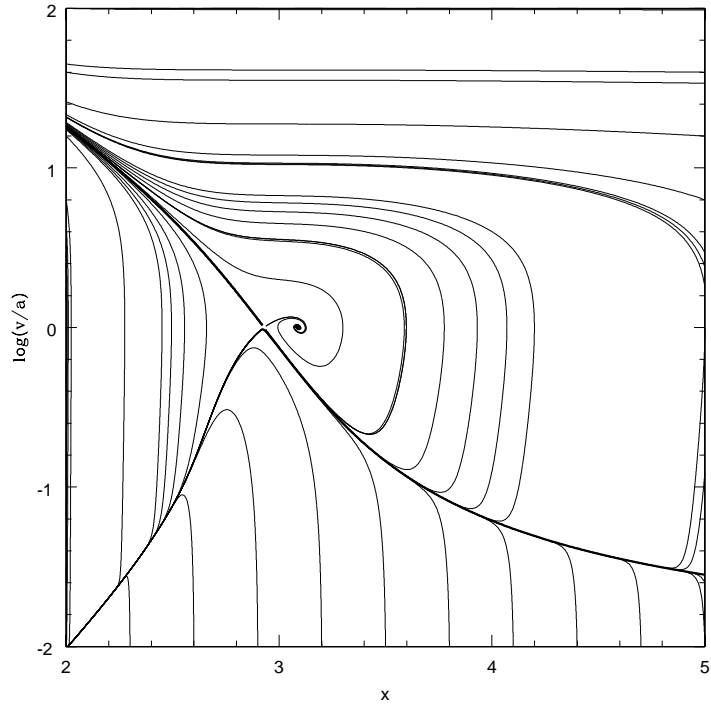}{8cm}{0}{50}{50}{-160}{-80}
\caption{The Phase Portrait for $a=0.01, \alpha = 0.04$ and $x_c\simeq 2.930$.
The physical solution (thick curve) passes a critical point of 
the saddle type.}
\end{figure}
\begin{figure}[t]
\plotfiddle{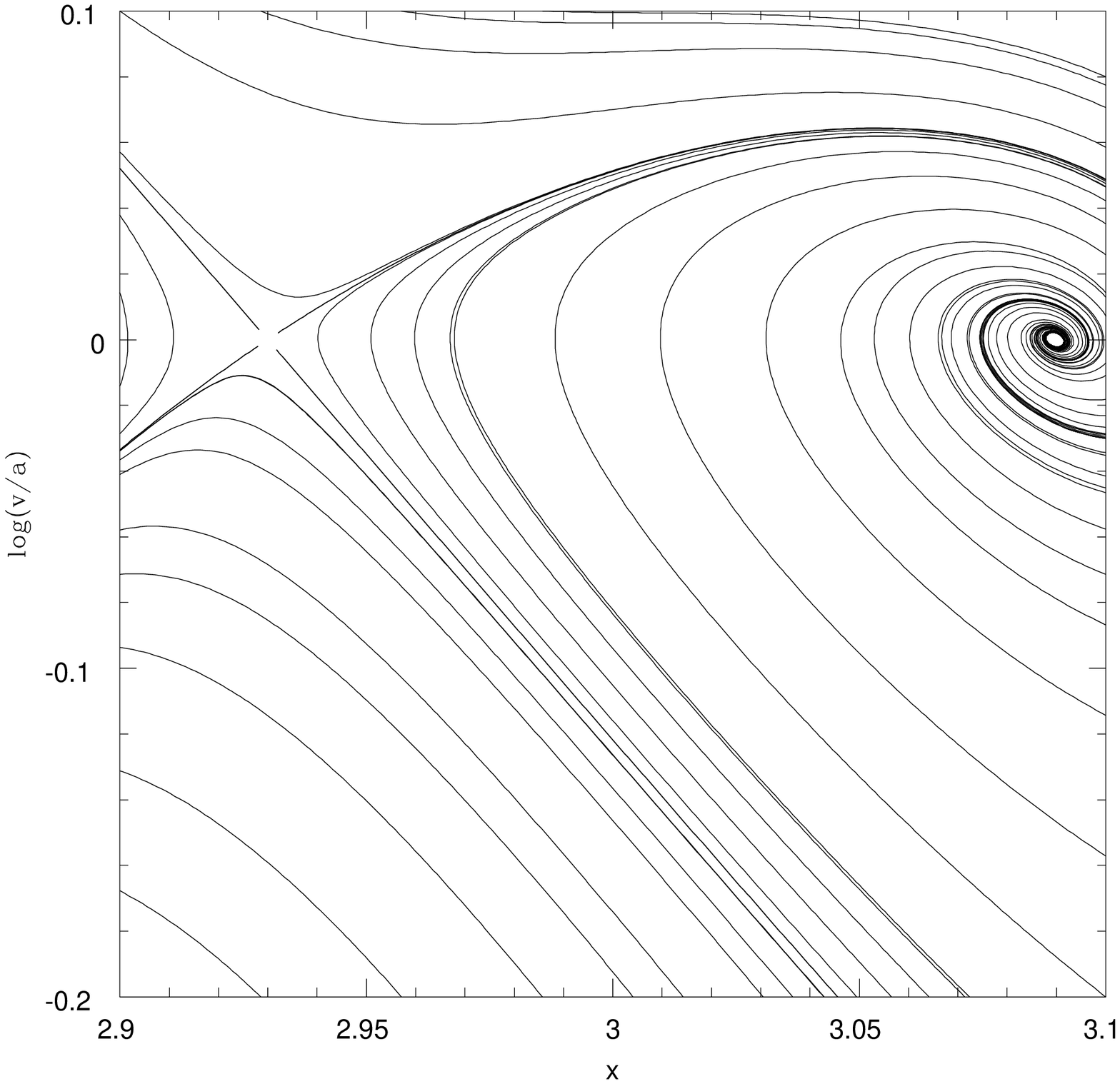}{8cm}{0}{50}{50}{-160}{-80}
\caption{The vicinity of the critical points in the phase portrait  
of Fig. 7.}
\end{figure}
\begin{figure}[t]
\plotfiddle{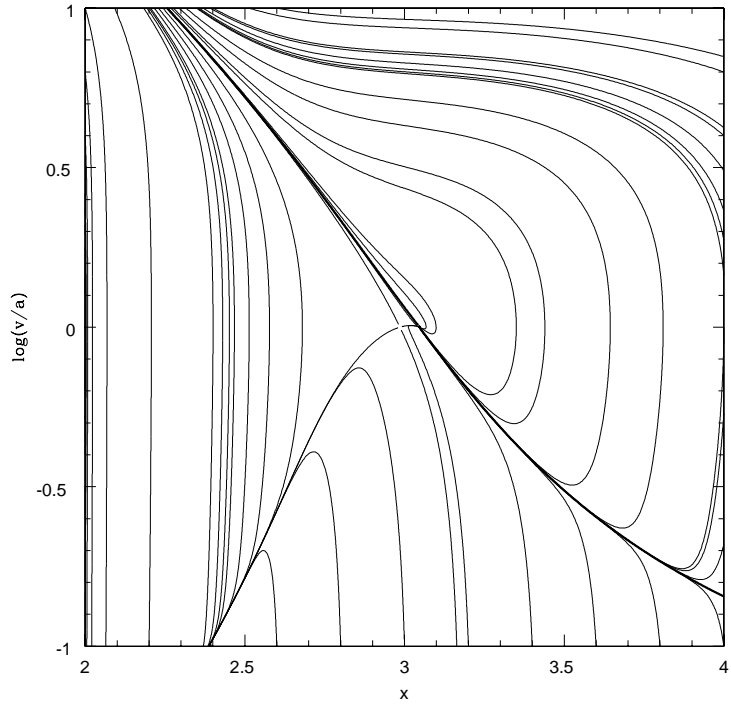}{8cm}{0}{50}{50}{-160}{-80}
\caption{The Phase Portrait for $a=0.01, \alpha = 0.1$ and $x_c\simeq 3.047$.
The physical solution (thick curve) passes a critical point of 
the nodal type, in the fast direction.}
\end{figure} 

\begin{figure}[t]
\plotfiddle{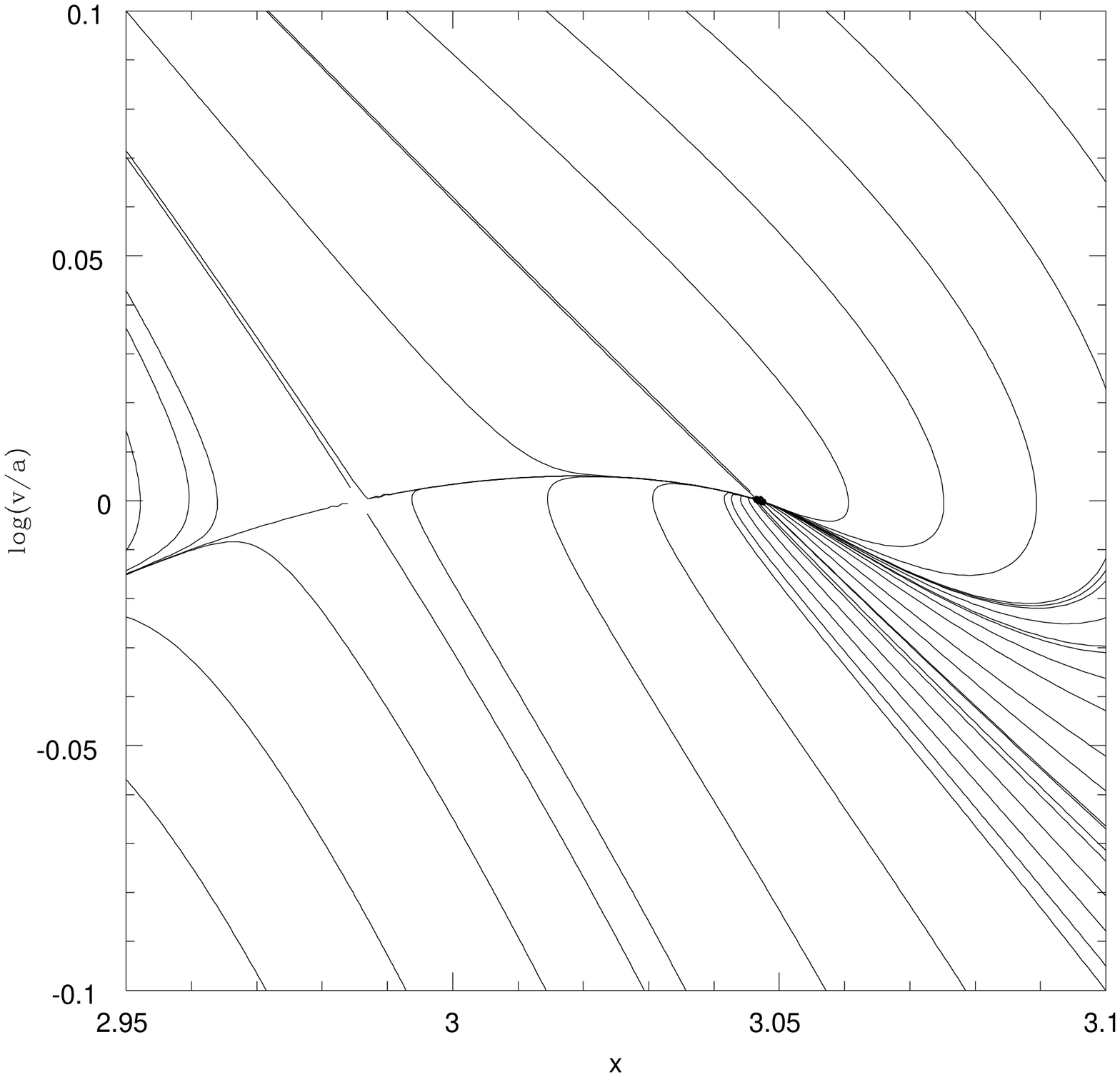}{8cm}{0}{50}{50}{-160}{-80}
\caption{The vicinity of the critical points in the phase portrait  
 of Fig.(9).}
\end{figure} 
\begin{figure}[t]
\plotfiddle{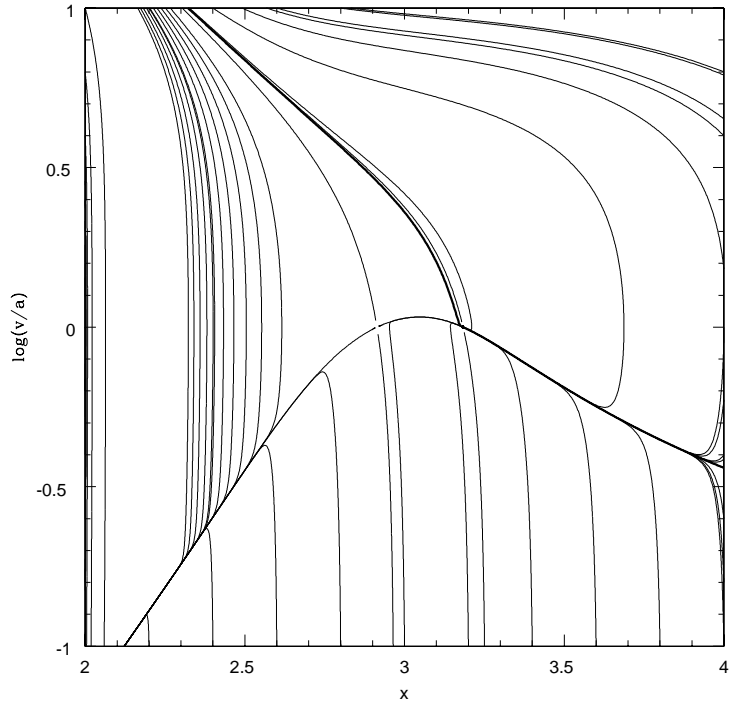}{8cm}{0}{50}{50}{-160}{-80}
\caption{The Phase Portrait for $a=0.01, \alpha = 0.3$ and $x_c\simeq 3.183$.
The physical solution (thick curve) passes a critical point of 
the nodal type, in the slow direction, analytically.}
\end{figure} 
\begin{figure}[t]
\plotfiddle{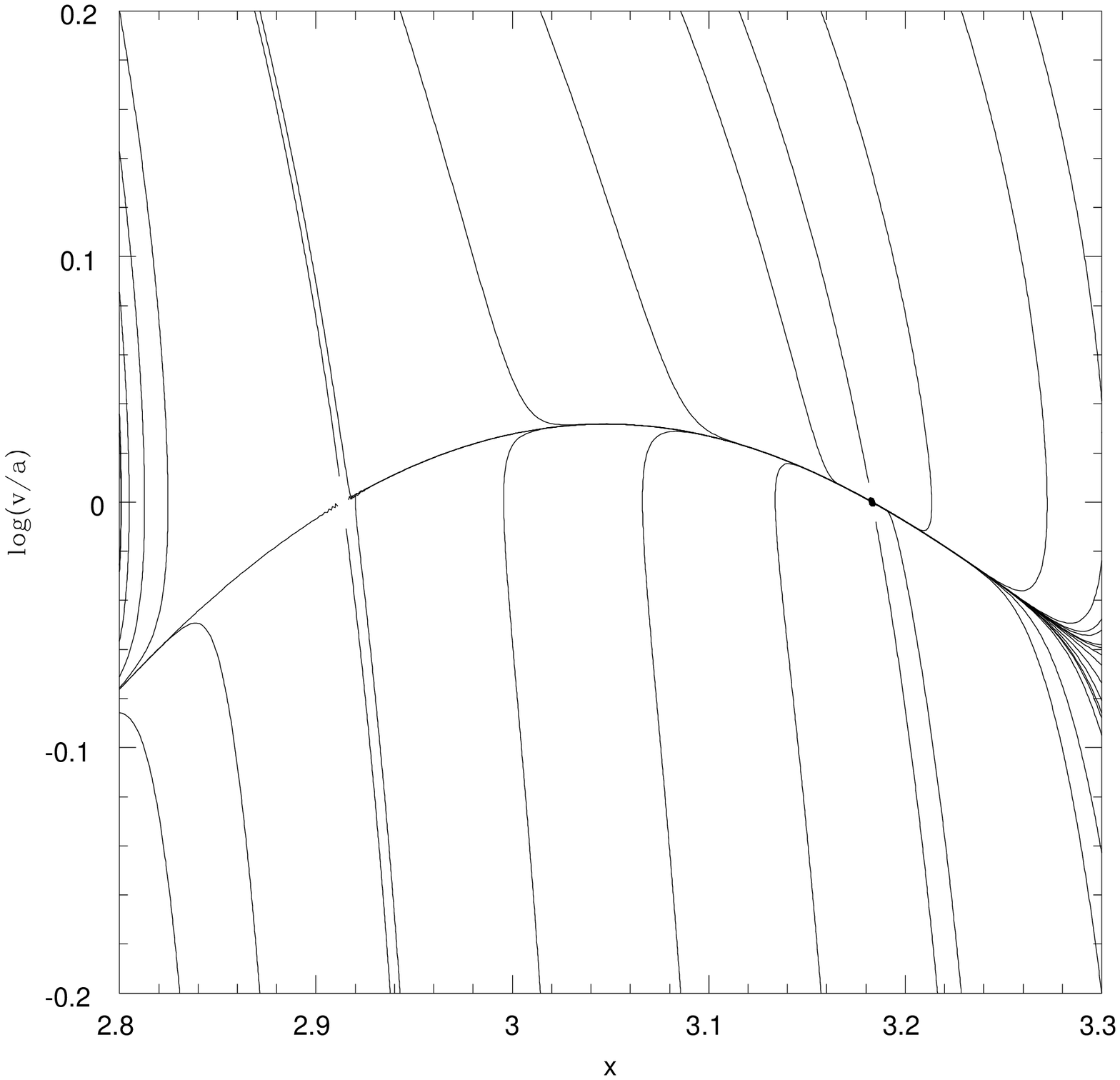}{8cm}{0}{50}{50}{-160}{-80}
\caption{The vicinity of the critical points in the phase portrait 
 of Fig.(11).}
\end{figure} 
\begin{figure}[t]
\plotfiddle{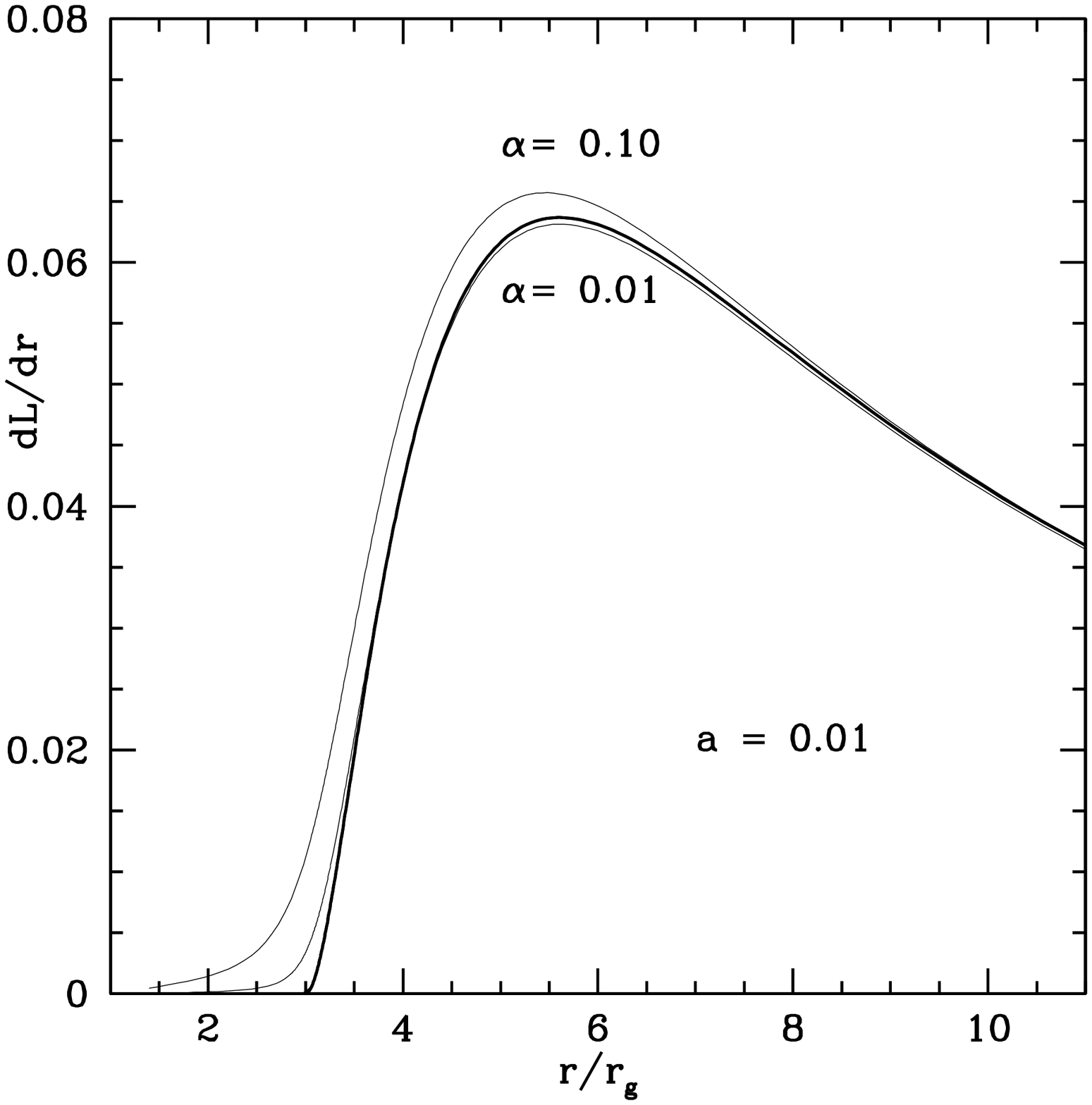}{8cm}{0}{50}{50}{-160}{-80}
\caption{The energy generation rate in the classical model is shown with
a thick solid line, and for our $ a = 0.01 $
disk model with two thin lines, corresponding
to $ \alpha = 0.01 $ and $ \alpha = 0.1 $, respectively.
The disk luminosity is normalized to unity for the classical model.}
\end{figure} 
\begin{figure}[t]
\plotfiddle{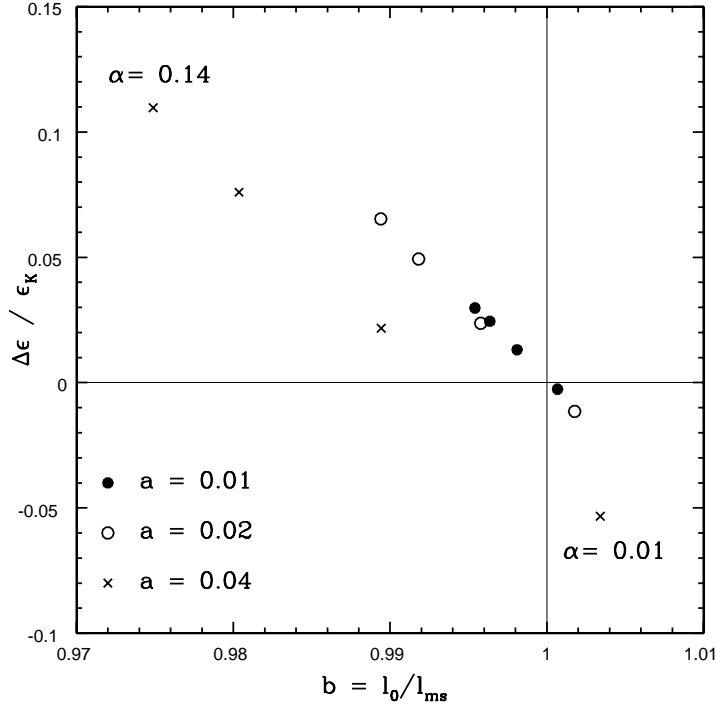}{8cm}{0}{50}{50}{-160}{-80}
\caption{
The relative change in the accretion efficiency for our disk models with
constant effective speed of sound: $ v_s = ac $ are shown for three values
of $ a $ (0.01, 0.02, 0.04) and for four values of the parameter $ \alpha $
(0.01, 0.05, 0.10, 0.14).  For $ \alpha = 0.01 $ the angular momentum
constant $ l_0 $ is smaller than $ l_{ms} $ and the accretion efficiency
is reduced compared to the model with exactly zero torque at $ r_{ms} $.
The efficiency of accretion increases with increasing $ \alpha $ and
with increasing disk thickness, which is proportional to $ a $.
}
\end{figure} 

\end{document}